\documentclass{article}%
\usepackage{amsmath}%
\usepackage{amsfonts}%
\usepackage{amssymb}%
\usepackage{graphicx}
\usepackage[a4paper,bindingoffset=0.2in,%
            left=1.5cm,right=1.5cm,top=2cm,bottom=1.5cm,%
            footskip=.25in]{geometry}
\usepackage{siunitx,subfig,xspace}
\usepackage{hyperref}
\newcommand*\aap{A\&A}

\newcommand*\aapr{A\&A~Rev.}
\newcommand*\aaps{A\&AS}

\newcommand*\apj{ApJ}
\newcommand*\apjl{ApJ}

\newcommand*\araa{ARA\&A}

\newcommand*\mnras{MNRAS}

\newcommand*\nat{Nature}

\begin{document}

\title{Positron annihilation signatures associated with the outburst of the microquasar V404 Cygni}
\author{Thomas Siegert$^1$\thanks{E-mail: \href{mailto:tsiegert@mpe.mpg.de}{tsiegert@mpe.mpg.de}}, Roland Diehl$^1$, Jochen Greiner$^1$, Martin G. H. Krause$^{1,2}$,\\
Andrei M. Beloborodov$^3$, Marion Cadolle Bel$^4$, Fabrizia Guglielmetti$^1$,\\
Jerome Rodriguez$^5$, Andrew W. Strong$^1$ and Xiaoling Zhang$^1$\\
\\
\tiny 1) Max Planck Institut f\"ur extraterrestrische Physik, D-85748 Garching, Germany\\
\tiny 2) Universit\"ats-Sternwarte M\"unchen, Ludwig-Maximilians-Universit\"at, Scheinerstra\ss e 1, 81679 M\"unchen, Germany\\
\tiny 3) Physics Department and Columbia Astrophysics Laboratory, Columbia University, 550 West 120th Street, New York, New York 10027, USA\\
\tiny 4) Max Planck Computing and Data Facility, Gießenbachstra\ss e 2, 85748 Garching, Germany\\
\tiny 5) Laboratoire Astrophysique Instrumentation Mod\'{e}lisation, UMR 7158, CEA/CNRS/Universit\'{e} Paris Diderot, CEA DSM/IRFU/SAp, 91191 Gif-sur-Yvette, France			
}

\date{Received 06 August 2015; Accepted 04 January 2016; Published online 29 February 2016\thanks{\href{http://www.nature.com/nature/journal/vaop/ncurrent/full/nature16978.html}{doi:10.1038/nature16978}}}
\maketitle

\begin{abstract}
Microquasars \cite{Mirabel1992_511,Mirabel1999_jets,Fender2004_xrb,Mirabel1994_jets} are stellar-mass black holes accreting matter from a companion star \cite{Remillard2006_xrb} and ejecting plasma jets at almost the speed of light. They are analogues of quasars that contain supermassive black holes of $10^6$ to $10^{10}$ solar masses. Accretion in microquasars varies on much shorter timescales than in quasars and occasionally produces exceptionally bright X-ray flares \cite{Greiner1996_GRS1915}. How the flares are produced is unclear, as is the mechanism for launching the relativistic jets and their composition. An emission line near 511 kiloelectronvolts has long been sought in the emission spectrum of microquasars as evidence for the expected electron-positron plasma. Transient high-energy spectral features have been reported in two objects \cite{Bouchet1991_511,Goldwurm1992_novamusca}, but their positron interpretation \cite{Sunyaev1992_xrb511} remains contentious. Here we report observations of $\gamma$-ray emission from the microquasar V404 Cygni during a recent period of strong flaring activity \cite{Kuulkers2015_V404}. The emission spectrum around 511 kiloelectronvolts shows clear signatures of variable positron annihilation, which implies a high rate of positron production. This supports the earlier conjecture that microquasars may be the main sources of the electron-positron plasma responsible for the bright diffuse emission of annihilation $\gamma$-rays in the bulge region of our Galaxy \cite{Prantzos2011_511}. Additionally, microquasars could be the origin of the observed megaelectronvolt continuum excess in the inner Galaxy.
\end{abstract}

\section{Main article}

Flaring activity of V404 Cygni was discovered with the Swift/BAT and MAXI X-ray monitors \cite{Kuulkers2015_V404}, and observations with INTEGRAL \cite{Winkler2003_INTEGRAL} started within two days, lasting from 17 to 30 June 2015. This recent activity period of V404 Cygni - 11 days (15-26 June) - was shorter than its previous outbursts \cite{Rodriguez2015_V404} in 1934 and 1989. The source exhibited multiple flares within hours, outshining \cite{Rodriguez2015_V404} the brightest persistent X-ray source in the sky, the Crab nebula, by factors of up to 40. V404 Cygni represents a ‘gold standard’ for multi-wavelength observations, as the parameters of the binary system are well known. It is composed of a 9-solar-mass black hole with a companion star of 0.7 solar masses \cite{Khargharia2010_V404} in a 6.5-day orbit \cite{Miller-Jones2009_V404} and a 67° inclination angle \cite{Khargharia2010_V404} to the line of sight, and is located at a distance of 2.4~kpc \cite{Casares1992_V404}.

INTEGRAL/SPI spectrometer data were extracted and calibrated following the standard procedures, including careful accounting for the detector response and background (see Methods section ‘SPI data extraction’). For spectral analysis, the data have been summed into three epochs of $\sim3$ days during the V404 Cygni flaring period. In each of these three epochs (INTEGRAL orbits 1554, 1555 and 1557) we have detected a significant excess of emission around 511~keV, consistent with positron annihilation.

At energies below 200~keV, the spectrum is well described by the standard model \cite{Done2007_xrb} of thermal Comptonization plus reflection, but at higher energies a large excess ($\sim18$~s.d.) appears above its high-energy tail (see Fig. 1). To further quantify this excess, we added a model spectrum of electron-positron plasma with temperature $T$ (see Methods section ‘Spectral fitting’, Extended Data Fig. 1 and Extended Data Table 1); the temperature serves as a measure of the annihilation line width.

\begin{figure}[t]
	\begin{center}
	\includegraphics[width=0.5\columnwidth]{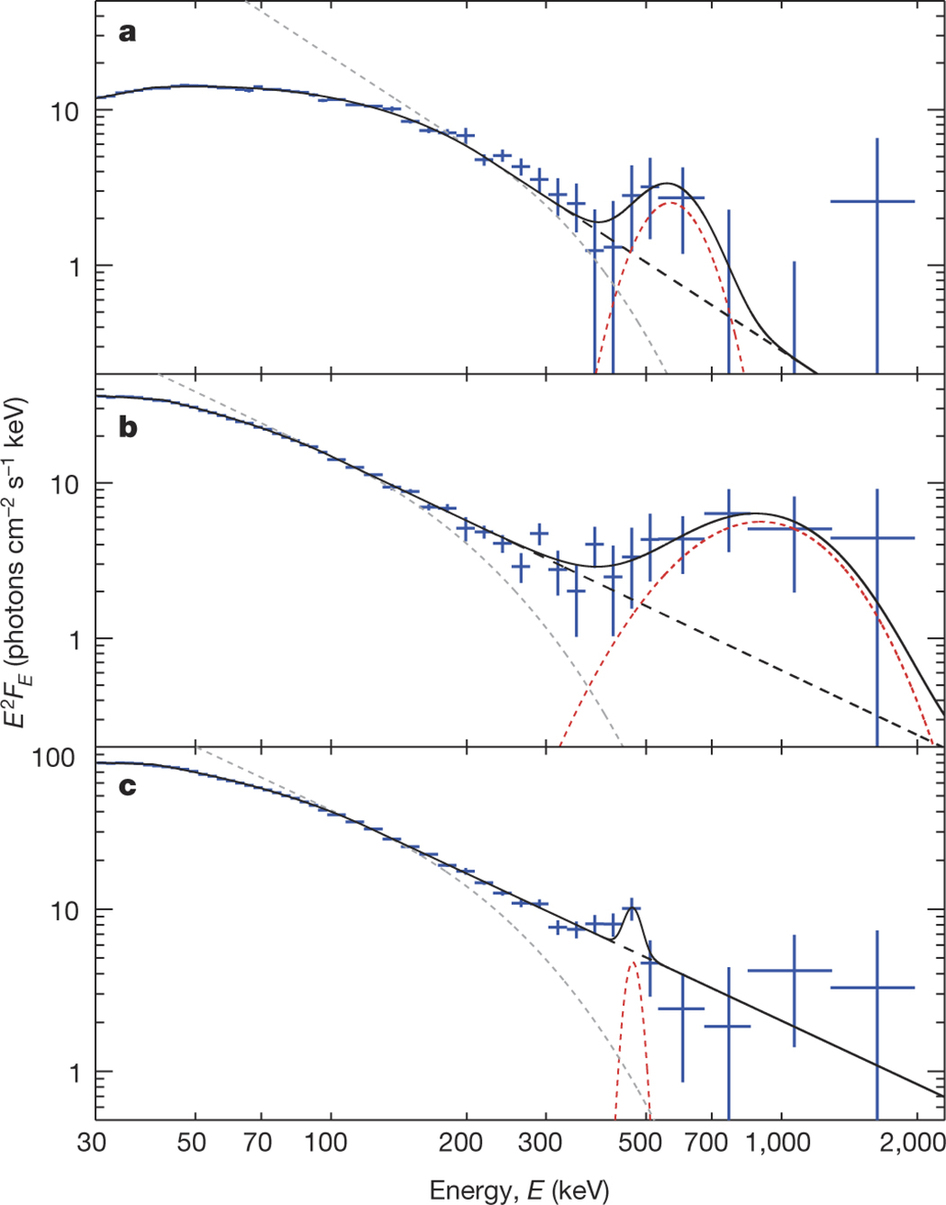}%
	\caption{Spectral evolution of V404 Cygni. a-c, Spectra in the soft $\gamma$-ray band in three different flaring epochs (a-c show the spectra measured in INTEGRAL orbits 1554, 1555 and 1557, corresponding to epochs 1, 2 and 3, respectively). Data (blue; error bars, 1~s.d.) are fitted as the sum of the Comptonization continuum (black dashed curve) and annihilation radiation from a relativistic hot plasma (red dashed curve). The standard thermal Comptonization model (grey dashed curve) fits the data up to $\sim200$~keV; it declines exponentially at higher energies and falls short of the observed flux. Our conservative, modified continuum model follows a power law instead. $E$, energy; $F_E$, flux at energy $E$.}%
	\label{fig:fig1}%
	\end{center}
\end{figure}

The characteristic curved spectral shape of pair annihilation emission describes the excess in all three epochs well. Its inclusion leads to a 5-s.d. overall improvement of the fit compared to a model describing the excess with a power law (see Extended Data Tables 2, 3, 4 for details, including model alternatives). No shift with respect to the laboratory frame of the annihilation line is required to fit the data in epochs 1 and 2. A redshift of $\sim$10\% is required in epoch 3. The line width varies strongly between the three epochs—the temperature parameter varies from a few keV to about 200~keV.

Electron-positron pair production is expected near luminous accreting black holes when their radiation spectra extend above $E = m_e c^2 = 511~\mathrm{keV}$ \cite{Maciolek-Niedzwiecki1995_xrb511,Svensson1987_xrb511}, where $m_e$ is the mass of the electron, and $c$ is the speed of light. Pairs are produced in collisions between MeV $\gamma$-rays, with an average cross-section $\sigma_{\gamma\gamma} \approx 10^{-25}~\mathrm{cm^2}$. This process is efficient because of the small size of the source, radius $r \approx (3-10) r_g$, where $r_g \approx 10~\mathrm{km}$ is the gravitational radius of the black hole. During the V404 Cygni flares, the observed luminosity in photons of energy $E \approx m_e c^2$ increases up to $L_1 \approx 10^{37}~\mathrm{erg~s^{-1}}$, which corresponds to a photon density $n_1 \approx L_1 / (\pi r^2 m_e c^3)$, and an optical depth for collisions between soft $\gamma$-rays of $\tau_{\gamma\gamma} \approx n_1 \sigma_{\gamma\gamma} r$. Gamma-rays of higher energies are absorbed more efficiently as they can interact with the more numerous X-rays of lower energies. The large optical depth $\tau_{\gamma\gamma} \approx 1$ tends to suppress the spectrum of the central compact source at photon energies $E \gg m_e c^2$, consistent with the upper limit on the GeV emission (obtained from analysing public Fermi/LAT data, $F_{GeV} < 10^{-6}~\mathrm{photons~cm^{-2}~s^{-1}}$; see Methods section ‘FERMI/LAT data analysis’). Thus, a significant fraction of the luminosity is converted to pair plasma: the plasma is continually created and annihilated inside the source, forming a broad annihilation line of width equivalent to $kT \approx 100~\mathrm{keV}$ \cite{Svensson1987_xrb511}. Photon collisions outside the source create a pair outflow \cite{Beloborodov1999_511}. Pairs annihilating outside the source are in Compton equilibrium with radiation at a temperature of a few tens of keV. Pair outflows have also been invoked to explain the shape of the X-ray spectra of accreting black holes, in particular their flat slopes and the reduced reflection component \cite{Fabian2015_AGN,Beloborodov1999_cygx1}.

These expectations are confirmed by our observations of a broad annihilation feature in epochs 1 and 2. The observed flux corresponds to a positron creation rate of $\dot{N_{\pm}} \approx 10^{42}~\mathrm{s^{-1}}$, and to an energy generation rate in the form of pairs, $L_{\pm}$, of $L_{\pm} \approx \dot{N_{\pm}} m_e c^2 \approx 10^{36}~\mathrm{erg~s^{-1}}$, a few per cent of the source luminosity \cite{Rodriguez2015_V404}, in agreement with models \cite{Beloborodov1999_511}. Pair plasma annihilating outside the source is then consistent with observations in epoch 1 ($kT \approx 30~\mathrm{keV}$), while annihilation inside the source corresponds to the broad line observed in epoch 2 ($kT \approx 170~\mathrm{keV}$).

In epoch 3, there is a hint of a steep decline on the blue side of the line that is characteristic of three-photon annihilation of positronium atoms (see Methods section ‘Spectral fitting’). Figure 2 shows the data fitted by a redshifted, narrow annihilation feature (Fig. 2a), or alternatively by a positronium-annihilation spectral shape (Fig. 2b). In addition, the broad component extending above 511~keV may be fitted by a very hot annihilation line; this more complex model would correspond to two annihilation regions, cold and hot (see Extended Data Table 5).

\begin{figure}[t]
	\begin{center}
	\includegraphics[width=0.55\columnwidth]{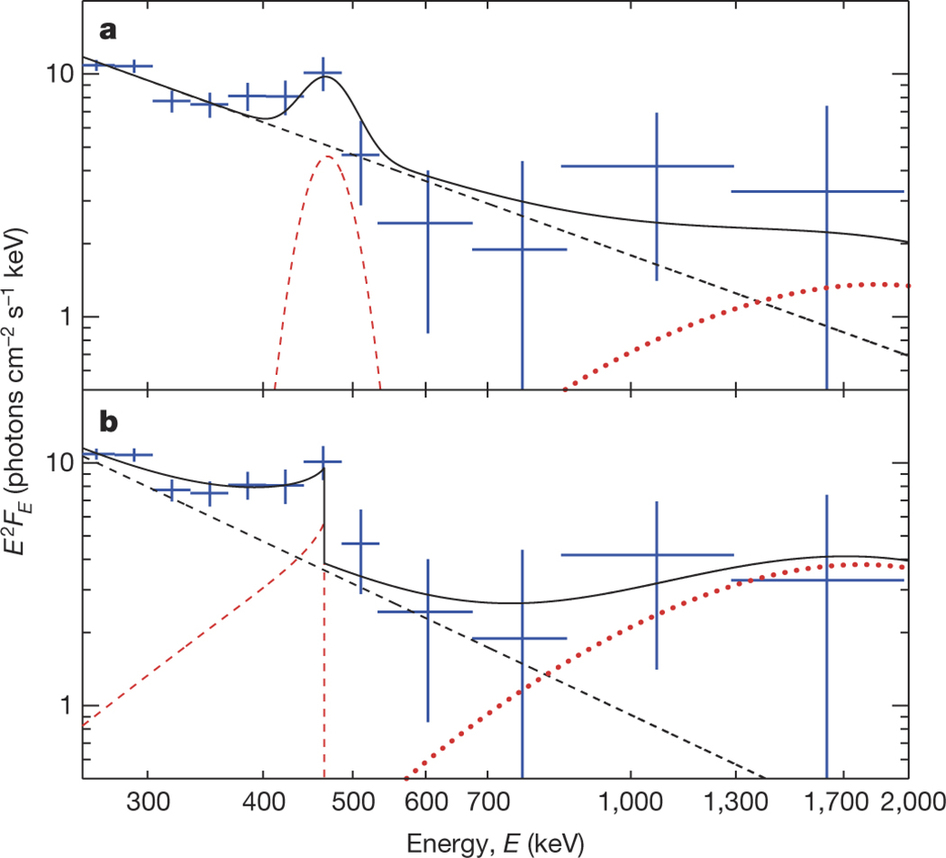}%
	\caption{Model alternatives for epoch 3. Shown is the high-energy part of the spectrum in epoch 3 fitted with different annihilation models. Data are shown blue in both panels (error bars, 1~s.d.). In both panels, the black dashed curve shows the Comptonization continuum, and the black solid line, the total model spectrum. a, Two components of annihilation in flight in a thermal plasma: $kT \approx 4~\mathrm{keV}$ redshifted by $\sim$10\% (red dashed curve), and $kT \approx 500~\mathrm{keV}$ (red dotted curve). b, Positronium three-photon annihilation redshifted by $\sim$10\% (red dashed curve) and annihilation in flight in a hot plasma (red dotted curve).}%
	\label{fig:fig2}%
	\end{center}
\end{figure}
\vspace{-10pt}

The smaller width of the line in epoch 3 ($kT \approx 2-3~\mathrm{keV}$), in combination with the detected redshift, poses a challenge to models. The 10\% redshift is consistent with a gravitational redshift of the source at radius $r \approx 10 r_g$. A line emitted in this region is expected to be significantly broader, owing to the mildly relativistic bulk motions of the plasma and dispersion in gravitational redshift.

The positronium interpretation also poses problems, because it requires \cite{Crannell1976_511} a dense plasma with a low temperature, $T < 10^6$~K. The minimum temperature of a source with luminosity $L \approx 10^{38}~\mathrm{erg~s^{-1}}$ and size $r \approx 10 r_g \approx 100$~km is its effective blackbody temperature, $T_{eff} \approx [L/(4 \pi r^2 \sigma_{bb})]^{1/4} \approx 10^7$~K, where $\sigma_{bb}$ is the Stefan-Boltzmann constant. The accretion flow at larger distances may contain much cooler gas, especially if the central radiation is beamed and shielded by the inner accretion disk. Then the pair plasma created in the central source may be blown out by radiation along the magnetic field lines towards the colder gas, decelerate there and form positronium before annihilating. This scenario is, however, inconsistent with the redshift of the line, in particular when interpreted as a gravitational redshift.

The morphology of the material near the black hole may also be inferred from soft X-ray observations at high spectral resolution \cite{King2015_V404}, suggesting high ionization out to $r \approx 2 \times 10^6$~km. This is a plausible value for the outer edge of the accretion disk. Variability of X-ray line strengths and a changing ionization strongly suggest that most of the X-rays are not received directly but are observed after reflection/reprocessing in the outer region. This is consistent with the flat shape of the hard X-ray continuum \cite{Natalucci2015_V404} up to 100~keV. In this picture, the outer edge of the disk is elevated, blocking our direct view towards the inner disk, and the X-ray continuum variability could be due to changes in the outer disk, rather than to rapid changes in the accretion rate onto the black hole.

The flux evolution of the continuum and annihilation features throughout our observation period is shown in Fig. 3. The two components appear to be correlated. When the X-ray flaring of V404 Cygni faded, the annihilation signal also vanished. However, some annihilation radiation appeared between the X-ray flares. Our time resolution is much longer than the dynamical timescale of the inner accretion disk (milliseconds), the jet ejection timescale \cite{Mirabel1999_jets} (minutes), the light travel time between the two binary components (80~s), and the timescale of adiabatic expansion of ejected blobs \cite{Fender2004_xrb} (minutes, less than 1~h). Thus, our measurements probe time-averaged values and do not resolve the rise time of the annihilation flux, which may be much shorter than our 6-h bins (see Methods section ‘Timing analysis’). Our search for a lag between the X-ray flares and the annihilation radiation was inconclusive, although there is an indication of a lag of several hours (see Methods section ‘Timing analysis’).

\begin{figure}[t]
	\begin{center}
	\includegraphics[width=0.75\columnwidth]{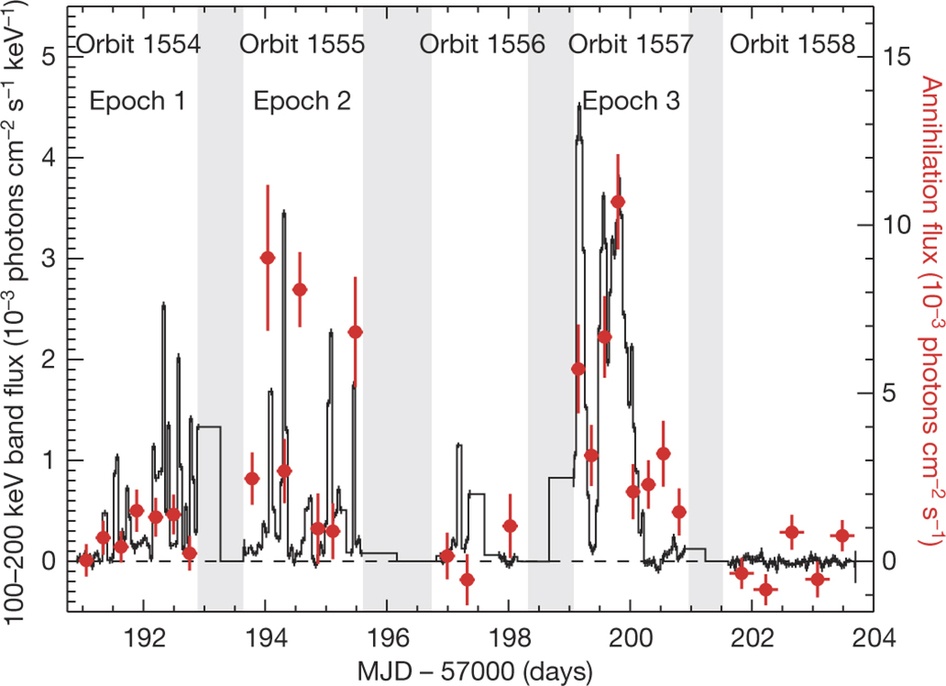}%
	\caption{Time history of X-ray continuum and annihilation emission. The two components of high-energy emission from V404 Cygni are shown as they evolve during the June 2015 flaring period, measured between 17 June and 30 June. INTEGRAL orbits 1554-1558 and epochs 1-3 are shown at the top. The black histogram shows the spectral photon flux in the 100-200~keV band (left-hand y axis) and represents the brightness of the Comptonization component emission. The red circles show the photon flux in the annihilation line averaged over 6~h (right-hand y axis); vertical error bars, 1~s.d. MJD, modified Julian date. Grey shaded areas mark the regions where no data have been taken.}%
	\label{fig:fig3}%
	\end{center}
\end{figure}

In contrast to hadronic gas, pair plasma is easily accelerated and immediately attains an equilibrium bulk speed $v \approx c/2$ away from the source (governed by the local radiation field anisotropy), forming the base of a relativistic outflow from the accretion disk \cite{Beloborodov1999_511}. The 67$^\circ$ inclination implies a Doppler blueshift of $\sim$10\%, comparable to the gravitational redshift at $r \approx 10 r_g$. From our data in epochs 1 and 2, we can neither confirm nor exclude possible residual shifts of the broad line. Annihilation radiation around the source can be strongly affected by the magnetic field configuration, which can change the outflow speed and direction. The power deposited in the pair plasma alone of $L_{\pm} \approx 10^{36}~\mathrm{erg~s^{-1}}$ would be sufficient to explain the observed typical radio luminosity associated with escaping blobs on larger scales \cite{Gallo2003_xrb}. A changing magnetic field may be responsible for the evolution of the line shape between epochs 1 and 2.

The outflow of pairs from the central source makes microquasars efficient factories, enriching the surrounding medium with positrons. This can help to solve the puzzle of annihilation radiation observed from the bulge region of our Galaxy \cite{Prantzos2011_511}. In a steady state, the annihilation rate in the inner Galaxy is $(1-2) \times 10^{43}~\mathrm{positrons~s^{-1}}$ and must be balanced by a positron supply with the same rate. During the 10-day flaring period, V404 Cygni produced $\sim 10^{42}~\mathrm{positrons~s^{-1}}$. In more luminous microquasars such as GRS~1915+105, this number is expected to be even larger. Thus, the steady-state annihilation rate in the inner Galaxy requires only about ten such sources to be active at any time. With duty cycles (that is, flaring versus quiescent epochs) of the order of $10^{-3}$, as observed for V404 Cygni, $\sim 10^3 - 10^4$ accreting black hole binaries of this type would be required in the inner Galaxy, consistent with population synthesis estimates \cite{Sadowski2008_xrb}.

The excess emission at a few hundred keV to a few MeV that we observed in the flaring epochs of V404 Cygni may also be responsible for the excess (about $10~\mathrm{keV~photons~cm^{-2}~s^{-1}}$ at 1~MeV \cite{Grenier2015_CR}) found in the diffuse $\gamma$-ray continuum spectrum of our Galaxy’s ridge \cite{Strong2005_gammaconti}. This excess is not accounted for by models of cosmic-ray interactions with interstellar gas \cite{Grenier2015_CR}, nor by inverse-Compton emission, and had been attributed to an unknown source population. A few active sources at an average observed microquasar distance of 5~kpc \cite{Gallo2003_xrb} would be needed to explain the diffuse MeV excess. The implied duty cycle is consistent with the above estimate for $\sim 10^4$ sources. Thus, the same population of microquasars could also be the origin of the observed inner Galaxy’s excess in MeV continuum.

\bibliographystyle{unsrt}


\appendix

\section{Methods: SPI data extraction}

INTEGRAL’s \cite{Winkler2003_INTEGRAL} SPI $\gamma$-ray spectrometer\cite{Vedrenne2003_SPI,Roques2003_SPI} provides photon event measurements, where pulse heights are recorded in a 19-element germanium semiconductor detector array. A coded mask imprints a shadowgram of celestial sources onto the detector array, as particular regions of the sky are blocked from detectors through the opaque tungsten elements of the mask. Re-orientations of the telescope by small angles (‘dithering’, $\sim 2^\circ$) provide an additional coding pattern. Coded-mask imaging allows sources within the telescope field of view of $\sim 30^\circ$ to be resolved with a precision of $\sim 2^\circ$.

We accumulate event data for each telescope pointing and detector and bin them into spectra, after initial preprocessing with INTEGRAL standard OSA~9~software. We use single-detector hits only, to ensure a well-known spectral detector response. Above 530~keV, we use pulse-shape filtered events to suppress an electronics malfunction that contaminates normal event data \cite{Weidenspointner2003_SPI}, and we apply their appropriate filter efficiency factor of 0.8. Pointings typically lasted one hour during the observations of V404 Cygni.

For adequate spectral precision, we accumulate three-day orbit data to extract a spectrum for V404 Cygni. The full set of detector spectra over an orbit is then fitted with parametrised models for instrumental background and for all candidate sources in the field of view \cite{Bouchet2008_gammasky}. The celestial emission models are folded into the data space of spectra using the coded mask shadowing properties, and the energy response of the detectors. Thus, intensity scaling factors per energy bin are obtained for all candidate sources within the telescope field of view, which are V404 Cyg, Cyg X-1, Cyg X-3, Cyg A, 3A1954+319 and EXO2030+375. Our model fit of background and sky contributions to data then provides spectra for all these sources.

The background is derived from data of the two preceding INTEGRAL orbits (1552, 1553). Being from the pre-flaring periods of V404 Cygni and being pointed to regions devoid of sources at high Galactic latitudes, these data are free of any high-energy ($> 300$~keV) signals from the sky. The background is modelled as a constant detector-intensity ratio (pattern) per energy bin, taken from these independent data. Owing to the strong time variability, the overall amplitudes per energy bin (background-intensity scaling factors) have been determined on a one-hour timescale.

The continuum sensitivity (3~s.d.) \cite{Roques2003_SPI} around 400~keV for an exposure of two days (corresponding to the dead-time corrected on-time of one orbit) is about $3.0 \times 10^{-6}~\mathrm{photons~cm^{-2}~s^{-1}~keV^{-1}}$ with only a weak dependence towards higher energies (about $1.8 \times 10^{-6}~\mathrm{photons~cm^{-2}~s^{-1}~keV^{-1}}$ at 1600~keV). This is sufficient to detect sources with intensities of $\sim 100$~mCrab above 400~keV. V404 Cygni, for example, shows an average intensity of $(12.5 \pm 3.5) \times 10^{-6}~\mathrm{photons~cm^{-2}~s^{-1}~keV^{-1}}$ around 500~keV, and about $(2.5 \pm 1.6) \times 10^{-6}~\mathrm{photons~cm^{-2}~s^{-1}~keV^{-1}}$1 around 1000~keV, during its flaring epochs (see Fig. 1), in accordance with sensitivity limits. Thus, at higher energies, statistical uncertainties dominate over systematics.

Some cross-talk among the sources is expected to occur in the data of a coded-mask instrument with a large field-of-view. Also, concerns about artefacts from background model inadequacies need special attention. We therefore compare the spectra of a set of hypothetical sources fitted additionally in the empty regions of the field-of-view, together with the above-mentioned known sources, as commonalities or anti-correlations among them would be a hint of background or cross-talk issues. For a given data set, the total number of measured photons is fixed, so that the inclusion of test sources may result in an increased flux value for one source while at the same time reduce the flux for another (anti-correlations). But apart from clear signals in the spectra of the real sources, all source spectra in the empty field regions are fully consistent with background only, and in particular do not show any spectral signatures beyond statistical fluctuations.

\section{Methods: Spectral fitting}

The spectra for the three flaring epochs were fitted as a sum of two components: a Comptonized continuum $C(E)$ and pair annihilation emission $P(E)$

\begin{equation}
f(E) = C(E) + P(E)
\label{eq:eq1}
\end{equation}

Continuum X-ray spectra of accreting black holes in the hard state are well explained by the thermal Comptonization model \cite{Done2007_xrb,Malzac2012_xrb}: seed low-energy photons (for example, thermal radiation from the optically thick accretion disk) are re-processed by a hot plasma corona through Compton scattering to higher energies, leading to a power-law spectrum. The power law cuts off exponentially where the photon energy exceeds the mean energy of the scattering electrons, which is typically $\sim 50-100$~keV \cite{Done2007_xrb,Malzac2012_xrb}. This model describes the spectra of V404 Cygni reasonably well below $\sim 200$~keV, between and during flaring \cite{Rodriguez2015_V404}. However, additional emission is required above 200~keV. One can model this emission as an additional power-law component due to scattering by non-thermal, accelerated particles. We use the following approximation to the Comptonization spectrum

\begin{equation}
C(E) =
   \begin{cases}
     A_0 \left( \frac{E}{E_0} \right)^{\alpha}                                             & \text{: } E < E_C \\
     A_0 \left( \frac{E}{E_0} \right)^{\alpha} \exp\left( \frac{E_C - E}{E_F} \right)      & \text{: } E_C \leq E \leq E_X \\
		 B_0 \left( \frac{E}{E_0} \right)^{\beta}                                              & \text{: } E > E_X
   \end{cases}
\label{eq:2}
\end{equation}

with $B_0 = A_0 \left( \frac{E_X}{E_0} \right)^{\alpha - \beta} \exp\left( \frac{E_C - E_X}{E_F} \right)$, where $E_0 = 100$~keV is a normalization convention.

The model of thermal pair annihilation (TPA) \cite{Svensson1987_xrb511,Svensson1996_tpa511}, $P(E)$, is calculated as follows. Electron and positron energy distributions are described by a Maxwell-J\"uttner distribution, which is the relativistic Maxwell-Boltzmann distribution

\begin{equation}
f(\gamma) = \frac{\gamma^2 \beta}{\Theta K_2(1/\Theta)} \exp\left( - \frac{\gamma}{\Theta} \right)
\label{eq:eq3}
\end{equation}

Here $\gamma = (1 - \beta^2)^{-1/2}$ is the Lorentz factor, $\beta = v/c$ is the dimensionless velocity, $\Theta = \frac{kT}{mc^2}$ is the dimensionless temperature of the pair plasma with Boltzmann’s constant $k$, electron (positron) mass $m$, the speed of light $c$, and $K_2$ is the modified Bessel function of second kind. The spectral distribution of annihilation photons can be calculated in terms of a dimensionless photon energy $x = \frac{h\nu}{mc^2}$ as

\begin{equation}
\frac{\text{d}n}{\text{d}t}(x,\Theta)\text{d}x = n_+ n_- c \text{d} x \frac{2}{\Theta K_2(1/\Theta)^2} \exp\left( -\frac{x}{\Theta} \right) \int_1^{\infty} \text{d}s 2(s-1) \sigma_{ann}(s) \exp\left( \frac{s}{x\Theta} \right)
\label{eq:4}
\end{equation}

Here, $n_{\pm}$ are the number densities of the positrons/electrons, $s=x_{cm}^2$ is the photon momentum in the centre-of-momentum frame, and $\sigma_{ann}(s)$ is the cross-section for the annihilation process. We have used simplified expressions for the temperature and energy regions of interest \cite{Svensson1996_tpa511} which are accurate to 0.04\% across the range presented in this analysis.

The TPA model ($P(E)$, equation (4)), shown in Extended Data Fig. 1, does not include possible Doppler shifts due to the bulk motion of the pair plasma or the gravitational redshift, which may be significant when the emission originates near the black hole.

The fitted parameters in our composite model, $f(E)$, are the normalization of continuum at 100~keV, that is, the amplitude $A_0$, the low-energy power-law index $\alpha$, the cutoff energy $E_C$, the e-folding energy $E_F$, the high-energy power-law index $\beta$, the extrapolation energy $E_X$, the amplitude of the annihilation feature $n_+ n_- c$, and the temperature of the pair plasma $T_{TPA}$. The parameters $E_0$ and $B_0$ do not influence the fit and are only introduced for convenience and clear arrangement.

From these parameters, we calculate the following derived parameters. $I_{TPA}$ is the differential flux integrated over the energy

\begin{equation}
I_{TPA} = \int_{-\infty}^{+\infty} P(E) \text{d}E
\label{eq:5}
\end{equation}

The plasma temperature TTPA is multiplied by the Boltzmann constant, $k$, for a conversion into keV units. Fitted and derived parameters for all epoch can be found in Extended Data Tables 1, 3, and 5.

The fit quality obtained for our one-orbit spectra is not satisfactory at first glance, with $\chi^2$ values of 63.7, 69.2 and 280.9 for 38 (37) d.o.f. (degrees of freedom) (see Extended Data Table 2). Typically, an observation is dominated by a large instrumental background. But below $\sim 200$~keV, the number of photons detected from V404 Cygni is very high, (about five times stronger than the background), thus systematics of the instrument response limit the accuracy of the fit. At energies above $\sim 150$~keV, our composite model obtains a satisfactory fit quality with $\chi^2$ values of 13.0 to 14.5 for 14 d.o.f. Overall, we detect a large high-energy excess above the conventional Comptonized cut-off description, with a total significance of $\sim 18$~s.d., consistent with high-energy excesses reported in other studies \cite{Rodriguez2015_V404,Natalucci2015_V404,Roques2015_V404}. Our more conservative analysis uses a phenomenological model of the continuum that avoids the high-energy cut-off of the Comptonization model by modifying it to a power-law extension towards high energies. When we use this modified Comptonization description, and add a model spectrum of electron-positron plasma of a temperature $T$, we find a significance of 5~s.d. altogether (see Extended Data Table 4) for the pair annihilation component above the Comptonization component. Significances have been estimated by $\chi^2$ goodness-of-fit tests, in which extra model components are zeroed in order to evaluate the improvement in $\chi^2$, compared to the simpler model with fewer parameters. The test statistics therefore follow a $\chi^2$-distribution with 1~d.o.f., in which data are properly scaled by their statistical uncertainties which account for Poissonian fluctuations of source and background.

The fit shown in Fig. 2b assumes a different annihilation model, more typical when positrons are slowed down by, for example, Coulomb interactions before annihilating \cite{Ore1949_511}. This has been well measured in our Galaxy \cite{Prantzos2011_511}, and in terrestrial laboratory experiments \cite{Sharma1978_511}. When positrons annihilate at thermal energies, the spectra of positron annihilation show the intermediate formation of a positronium atom, consisting of a positron and an electron. This process is very efficient at low energies and in particular below a threshold energy of 6.8~eV (equivalent to temperatures below $7.8\times10^4$~K). Positron annihilation then can occur from a singlet state of positronium with two photons at 511~keV (\textit{para}-positronium), or a triplet state with a three photon annihilation continuum spectrum \cite{Ore1949_511} rising in intensity up to a maximum at 511~keV (\textit{ortho}-positronium).

We define the fitted parameters of the additional components as follows. $E_{centroid}$ is the Doppler-shifted peak position of the low-energy annihilation feature, that is, pair plasma annihilation in Fig. 2a, and the sharp edge of the ortho-positronium shape, $O(E)$, in Fig. 2b. The plasma temperature $kT_{TPA}$ is only given for cases in which a thermal pair annihilation model has been fitted to the data. In both cases, the additional components are scaled/normalized by a fitted amplitude $O_0$.

We derive the flux, $F_{511}$, of the second features by integrating the differential spectral shape over the energy. For Fig. 2a, see equation (5), for the \textit{ortho}-positronium feature, the flux is given by

\begin{equation}
F_{511} = \int_{-\infty}^{+\infty} O(E) \text{d}E = O_0 (\pi^2 - 9)
\label{eq:}
\end{equation}

\section{Methods: Timing analysis}

For the timing analysis of the annihilation emission, we only fit the amplitudes of our model components, $A_0$ of the continuum, and $n_+ n_- c$ of the annihilation feature (see equations (2) and (4) and Fig. 3), in hourly time bins, while holding the spectral shape, that is, all other parameters in both continuum and annihilation model, fixed. This assumes a constant thermal pair annihilation temperature $T_{TPA}$ during one epoch and sums all emission which is spectroscopically identified as due to positron annihilation. The rise time of the excess flux is estimated as a factor of two increase in the positron annihilation component flux. This is difficult to estimate, as it is limited by photon statistics; we estimate a flux doubling within $\sim 2$~h. This interprets all fitted signatures on top of the continuum, $C(E)$, as pair plasma emission, rather than accelerated particle emission represented by a power law. In particular, above 500~keV, this may incur a bias towards high annihilation flux values, compared to the detailed spectroscopic analysis per epoch.

The time evolution of the hard X-ray continuum emission (100-200~keV) is obtained \cite{Rodriguez2015_V404} from fitting in one-hour intervals the coded-mask response to all sources in the field together with the background model, adjusting the background normalization coefficient per hour.

The linear Pearson correlation coefficient between the 100-200~keV band and the annihilation flux of V404 Cygni during MJD 57191 and 57203 in hourly time bins is 0.45 for zero lag, and 0.34 for a lag of $\sim 15$~h.

\section{Methods: FERMI/LAT data analysis}
We have analysed the all-sky survey data of the FERMI/LAT instrument taken during the 15.5-h time window when we see the largest positron flare, in epoch 3 at orbit 1557, corresponding to MJD 57199.616 to 57200.261. The Pass8 data of a 20$^\circ$ circle around V404 Cygni in the energy range 100~MeV to 10~GeV have been retrieved from the Fermi Science Support Center (\verb|http://fermi.gsfc.nasa.gov/cgi-bin/ssc/LAT/LATDataQuery.cgi|; accessed 6 August 2015). We employed the unbinned likelihood analysis as implemented in the user-contributed LATAnalysisScripts (\verb|http://fermi.gsfc.nasa.gov/ssc/data/analysis/scitools/LATAnalysisScripts.html|; accessed 6 August 2015) provided by FSSC, after changing to Pass8 details. Standard event cleaning/removal was applied (\verb|http://fermi.gsfc.nasa.gov/ssc/data/analysis/documentation/Cicerone/Cicerone_Data_Exploration| \\ \verb|/Data_preparation.html|; accessed 6 August 2015). We used the \verb|gll_iem_v06.fits| Galactic interstellar emission model, and the \verb|iso_P8R2_SOURCE_V6_v06.txt| isotropic spectral template for the SOURCE event class (front+back). The Cygnus Loop, Cygnus Cocoon, $\gamma$ Cyg and HB~21 have been included in the fitting, as well as bright 3FGL sources. No source is detected at the position of V404 Cygni. We derive upper limits of $8\times10^{-7}~\mathrm{cm^{-2}~s^{-1}}$ in the 100~MeV to 1~GeV band, and $3\times10^{-9}~\mathrm{cm^{-2}~s^{-1}}$ in the 1-10~GeV band.

\section*{Acknowledgements}
The INTEGRAL/SPI project has been completed under the responsibility and leadership of CNES; we are grateful to ASI, CEA, CNES, DLR, ESA, INTA, NASA and OSTC for support of this ESA space science mission. R.D. and J.G. are also supported by the Munich excellence cluster ‘Origin and evolution of the Universe’. M.G.H.K. is supported by the Deutsche Forschungsgemeinschaft, project number PR 569/10-1, as part of DFG Priority Program 1573. J.R. acknowledges funding support from the French Research National Agency, CHAOS project ANR-12-BS05-0009, and from the UnivEarthS Labex program of Sorbonne Paris Cit\'{e}.

\section*{Contributions}
T.S. was responsible for the spectroscopy analysis, data modelling, and paper writing, and R.D. led the analysis and paper writing. J.G., M.G.H.K. and A.M.B. were responsible for interpretational aspects and crucial inputs to the paper. J.G. was responsible for analysis of the Fermi data, M.C.B and J.R. contributed in microquasar physics, F.G. in data analysis, A.W.S. in $\gamma$-ray continuum and cosmic-ray physics, and X.Z. was responsible for data preparation and reduction, and the instrument response.

\section*{Extended Data Figures}

\begin{figure}[!ht]
	\begin{center}
	\includegraphics[width=0.75\columnwidth]{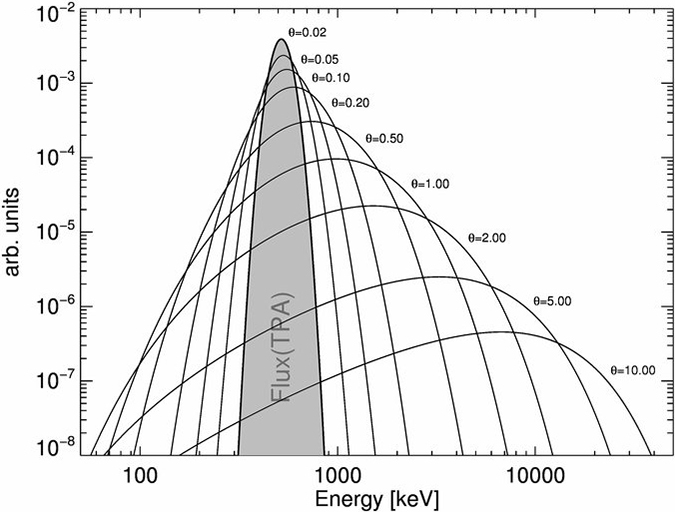}%
	\caption*{Extended Data Figure 1: Spectral shape of annihilation emission from a relativistic thermal pair plasma. Each curve shows intensity per unit energy and is labelled with $\Theta (= kT / m_e c^2)$, the dimensionless temperature. The model is used to quantify the width of the observed annihilation line (see Methods section ‘Spectral fitting’). The grey shaded area is the integrated flux over all energies for this thermal plasma annihilation model (TPA).}%
	\label{fig:ed_fig1}%
	\end{center}
\end{figure}

\section*{Extended Data Tables}

\begin{figure}[!ht]
	\begin{center}
	\includegraphics[width=0.75\columnwidth]{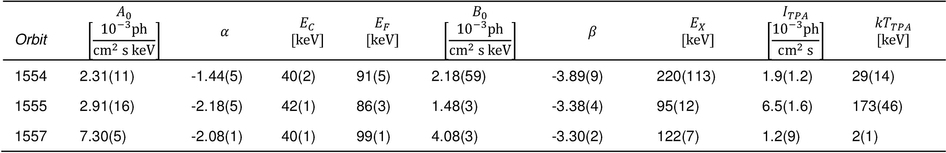}%
	\caption*{Extended Data Table 1: Spectral fit parameters for the flaring epochs of V404 Cygni. These epochs are shown in Fig. 1. Uncertainties are given in brackets, in units of the last digit. The redshift seen in orbit 1557 is 0.101(1). The amplitudes $A_0$ and $B_0$ are normalized to the flux at 100~keV. For nomenclature, see Methods section ‘Spectral fitting’.}%
	\label{fig:ed_tab1}%
	\end{center}
\end{figure}

\begin{figure}[!ht]
	\begin{center}
	\includegraphics[width=0.75\columnwidth]{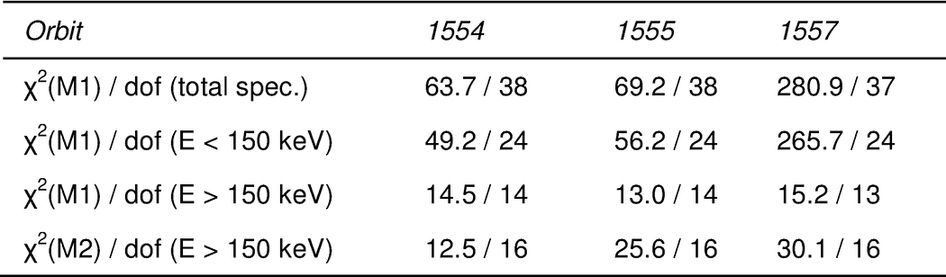}%
	\caption*{Extended Data Table 2: Goodness-of-fit for spectra. The goodness-of-fit for spectra corresponding to epochs shown in Fig. 1 is measured by $\chi^2$/d.o.f. (d.o.f., degrees of freedom). We illustrate systematics by giving results for different energy regions, and for different models (see Methods section ‘Spectral fitting’). Model 1 (M1) is the full thermal pair annihilation model, and model 2 (M2) is a high-energy cut-off power-law model for Comptonized disk emission plus another independent power law capturing accelerated particle emission.}%
	\label{fig:ed_tab2}%
	\end{center}
\end{figure}

\begin{figure}[!ht]
	\begin{center}
	\includegraphics[width=0.75\columnwidth]{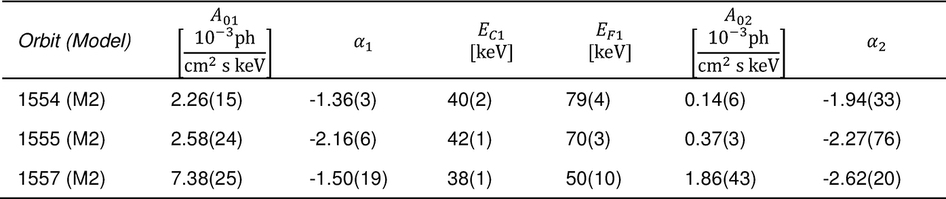}%
	\caption*{Extended Data Table 3: Spectral fit result details for the alternative model M2. See also Extended Data Table 2. Model 2 (M2) is as in Extended Data Table 2. The amplitudes $A_{01}$ and $A_{02}$ are normalized to the flux at 100~keV. For nomenclature, see Methods section ‘Spectral fitting’.}%
	\label{fig:ed_tab3}%
	\end{center}
\end{figure}

\begin{figure}[!ht]
	\begin{center}
	\includegraphics[width=0.75\columnwidth]{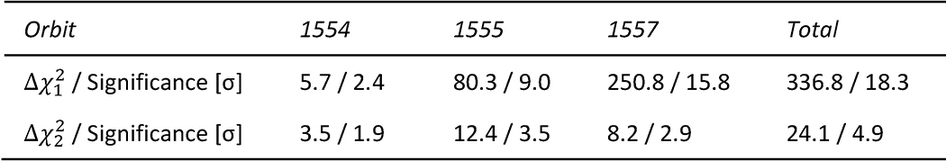}%
	\caption*{Extended Data Table 4: Significance estimates of additional components. $\Delta \chi_1^2$ describes the improvement of an additional high-energy extension over a high-energy cut-off power-law model only, and $\Delta \chi_2^2$ of an additional thermal pair annihilation feature over the model combining a high-energy cut-off power-law model and its high-energy power-law extension. Significances have been calculated by $\chi^2$-tests with one additional component each. For nomenclature, see Methods section ‘Spectral fitting’.}%
	\label{fig:ed_tab4}%
	\end{center}
\end{figure}

\begin{figure}[!ht]
	\begin{center}
	\includegraphics[width=0.75\columnwidth]{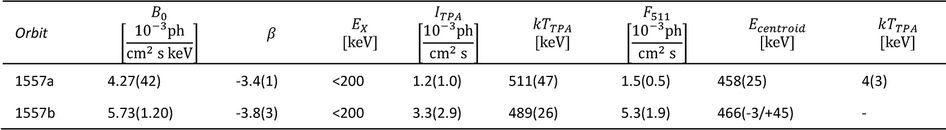}%
	\caption*{Extended Data Table 5: Spectral fit parameters for the third flaring epoch. The energy region between 200~keV and 2000~keV for orbit number 1557 is shown. In addition to a hot thermal annihilation plasma with temperatures around 500~keV, components are either another, Doppler-shifted, thermal annihilation line (1557a), or an\textit{ortho}-positronium continuum (1557b). The amplitude $B_0$ is normalized to the flux at~100 keV. The $\chi^2$ goodness-of-fit values are 9.3 and 5.9 for 9 and 8 d.o.f., respectively. For nomenclature, see Methods section ‘Spectral fitting’.}
	\label{fig:ed_tab5}%
	\end{center}
\end{figure}

\clearpage
\begin{center}
	\huge Separating Background and Sky in INTEGRAL/SPI Data \\
	\large Further Comments on Data Analysis - Applications to V404 Cygni \\
	\small Roland Diehl, Thomas Siegert, and Xiaoling Zhang \\
	\small \date{15 April 2016}
\end{center}
\section{Celestial and Background Signatures in SPI Data}
\label{sec:intro}

\begin{figure}[!ht]
  \centering
  \includegraphics[width=0.7\linewidth]{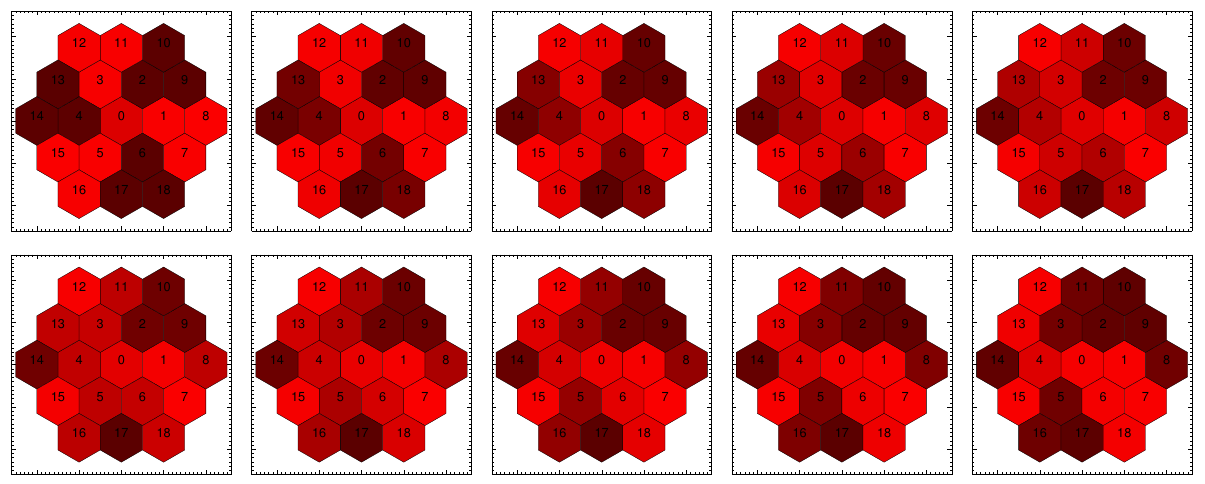}
   \caption{The 19 Ge detectors of SPI are arranged in a hexagonal, densely packed, configuration. Detectors are numbered inside-out, starting from 0 (central detector). The coded-mask shadowgram is shown here for ten pointings of the telescope in steps of 0.2 degrees (from Kretschmer, K. A., 2011).}
  \label{fig:SPI-maskpattern}
\end{figure}

SPI is a coded-mask telescope: The only information that distinguishes a set of photon events that arise from a celestial source from instrumental background are the detector patterns, i.e. the way the event counts are distributed over the 19 detectors of the camera during the time of the observation. The expectations of this pattern for a source shining through the mask of the SPI telescope are that some of the 19 detectors are \emph{shadowed} by the mask and thus do not record events from the celestial source, while others are exposed to the source (Fig.~\ref{fig:SPI-maskpattern}). There are in-between cases of partial shadowing, depending on the source aspect angle. The telescope is re-oriented per each $\sim$1/2 to 1 hour, an interval called a \emph{pointing}. The pattern of successive such pointings is normally a $5\times5$-grid around the target direction, with steps of $2.1^{\circ}$. This systematic variation of the telescope pointing around the target direction is called \emph{dithering}. Thus the mask shadow shifts from pointing to pointing, and in each successive case other detectors will be exposed to the source, or occulted.

The complex systematics of such exposure/occultation conditions determines the expected detector ratio patterns for a celestial source over the total exposure time, consisting of hundreds to many thousands of pointings. For source intensities of $\sim$10$^{-5}$~ph~cm$^{-2}$~s$^{-1}$ (see e.g. Diehl, 2013), the 255 cm$^2$ of typically exposed SPI detector area will capture only few photons per each telescope orientation, $(255/19)\times3600\times10^{-5}\sim 0.5$ events per detector for a one-hour pointing. Therefore this expected detector pattern is realised only in a statistical way through the number of totally detected source photons, and represents an \emph{a priori} probability per detector and pointing. The probability of observing $n$ detector counts per pointing from the source follows from Poissonian counting statistics. The product of those probabilities forms the likelihood of observing a source data set, given the source direction and intensity, and the pointing schedule of the SPI coded-mask instrument during the observation. 

On the other hand, the pattern of relative counts among detectors for what we believe is background is much more straightforward: Instrumental background hits detectors from all directions - the mask is not relevant here. The re-orientations by $2.1^{\circ}$ can be expected to not affect the relative exposures of detectors to instrumental background, which originates from the instrument itself and spacecraft materials. The exposure (more precisely the effective solid angle) of detectors for radiation from these materials remains constant, and a possible change in irradiation of the materials by cosmic ray particles will affect the total background intensity, but not the locations of their origins. Therefore, the detectors will see the same relative proportions of background counts throughout longer observations of the same sky region. Background spectra are dominated by a continuum of roughly power-law spectral shape, and lines are superimposed which may reach intensities of the underlying continuum for the stronger lines. These lines can be attributed to characteristic materials, as the atomic nuclei are activated by cosmic rays, and nuclear de-excitation lines are emitted. The spatial distribution of such a characteristic isotope within the instrument and spacecraft material determines how each of the 19 SPI Ge detectors will be irradiated. The pattern per each line energy thus will differ in general, but lines originating from the same isotope should have similar detector patterns. 

\begin{figure}[!ht]
  \centering
  \includegraphics[width=0.5\linewidth]{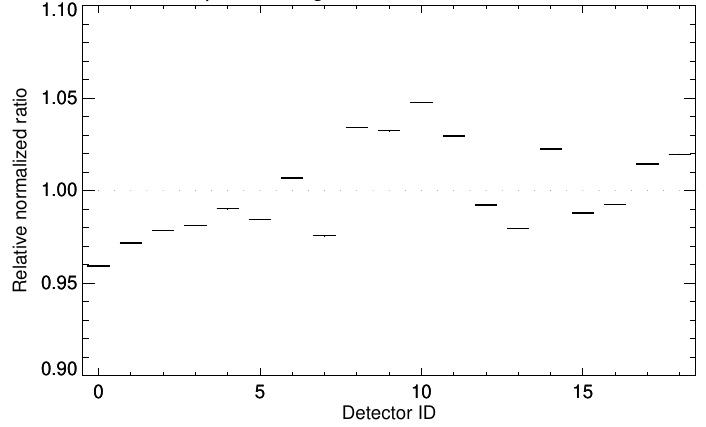}
  \includegraphics[width=0.5\linewidth]{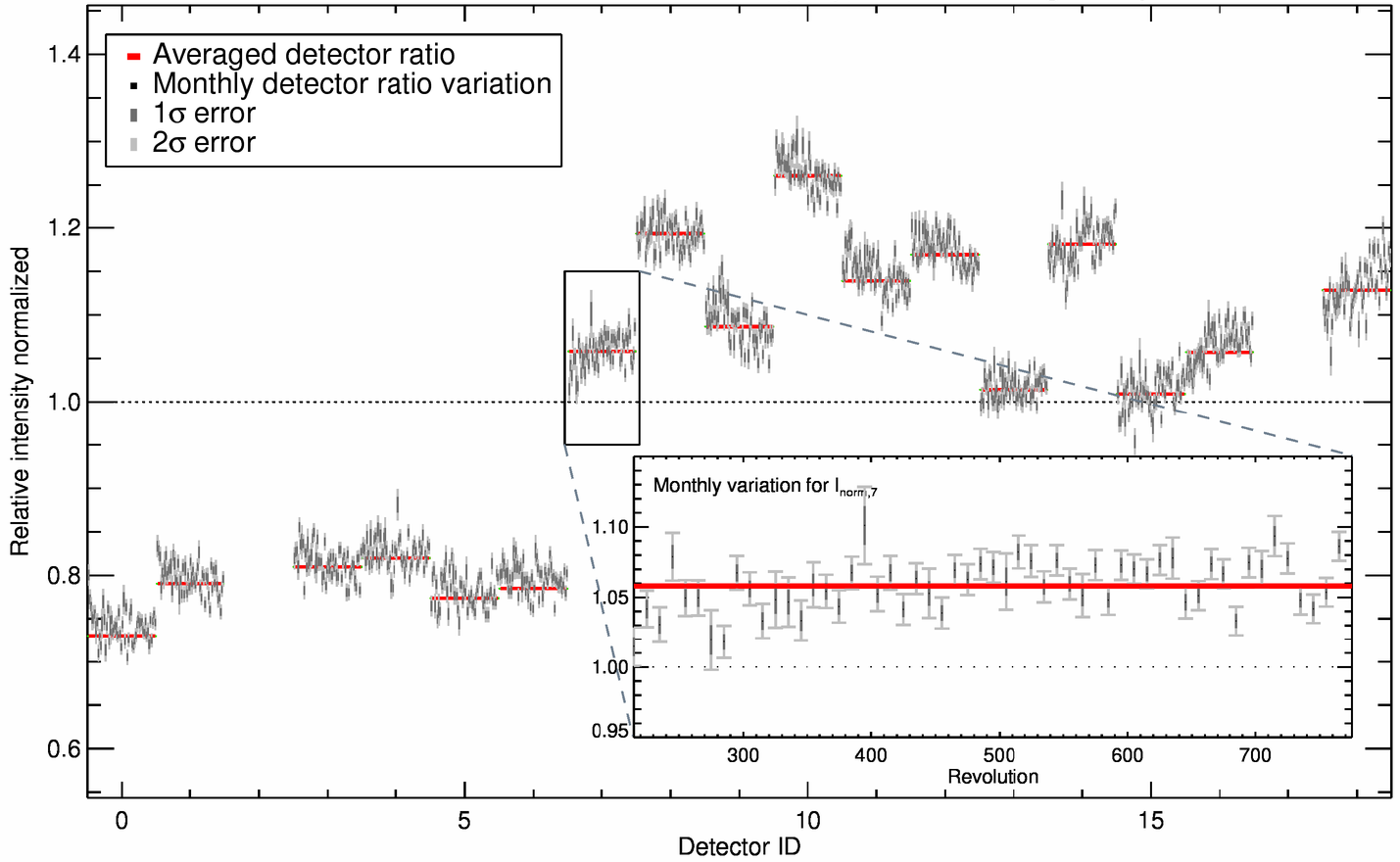}
   \caption{Ratio of intensities in different detectors (from Siegert, 2013). Top: The ratio among detectors varies only by a few percent in a broad continuum energy band, i.e. irradiation is rather homogeneous. Bottom: If specific line components are separated, the variation among detectors can be much larger, reflecting the specific origins of background in such lines. \emph{Insets:} The detector ratio for different times during the INTEGRAL mission shows statistical fluctuations, but the general characteristic detector pattern remains, and is rather stable.}
  \label{fig:SPI-detratios}
\end{figure}

This was confirmed in a detailed study (Siegert, 2013): In particular, lines originating from Ge and related isotopes (Ga, Zn, As, etc.) are more intense in detectors central to the camera and hence surrounded by Ge material on all sides, while detectors at the outskirts of the camera see less of this background, being neighboured by fewer Ge detectors. Conversely, lines that arise from the BGO material of the massive SPI anticoincidence system surrounding the detectors are more intense in outer Ge detectors, while detectors more central to the Ge camera are partially shielded from this background and thus record less intense backgrounds from Bi and related isotopes such as Pb or Ra (see Fig.~\ref{fig:SPI-detratios}). Those detector-to-detector variations can amount to more than 50\% in these extreme cases, but are smaller for lines from other materials. Continuum background also has its characteristic detector ratio signature (\emph{top graph} in Fig.~\ref{fig:SPI-detratios}), which may gradually change with energy. In general, the detector ratio for continuum is closer to one, each detector sees roughly equal intensities of continuum background, and over energy ranges below several tens of keV no significant variations occur (see \emph{inserts in bottom graph} of Fig.~\ref{fig:SPI-detratios}). The level of instrumental background depends on the intensity of the cosmic-ray irradiation of the spacecraft and instrument material. The sensitivity to such cosmic-ray bombardment may be  different for different lines and for the continuum.

Measuring celestial photon events against a large instrumental background thus is possible under three key assumptions: (1) The relative contributions of detectors to the source signal follows the mask shadowing. (2) The relative contributions of detectors to instrumental background can be determined from independent measurements, and normalisations to the observation data set on intermediate time intervals (typically one orbit) cater for variations that occur in time and energy. (3) Short-term intensity variations of the background can be traced through generic / integrated background level monitors, or determined from the data by normalisation; spectral or detector pattern remain constant across such intermediate time intervals.

\section{Measuring Celestial Sources Against Large Instrumental Background}
\label{sec:celestial-background}

With above three assumptions, in all applications of SPI, we measured celestial source signals that contribute only a small fraction of the total number of recorded events. SPI analysis procedures express measured data as 

\begin{equation}
\textbf{D} = d_{(i,j,k)} = \sum_{\alpha,\delta} {\Theta_i \cdot I_{\alpha,\delta;i} \cdot R_{\alpha,\delta;i,j,k}} + \sum_m {\Phi_{m,i,t} \cdot B_{m;i,j,k}} 
\label{eq:datamodel}
\end{equation}

with $i$ the identifier of an energy bin, $j$ the identifier of a detector, and $k$ the identifier of a telescope pointing, so that $d_{i,j,k}$ is the data content of bin $(i,j,k)$. $I_{\alpha,\delta;i}$ is the intensity of a celestial source at sky location ${\alpha,\delta}$ in energy bin $i$. $B_m$ is the background model component $m$. $R_{\alpha,\delta;i,j,k}$ is the instrumental response, i.e. the probability to measure counts in data bin $(i,j,k)$ from a celestial source located in sky region ${\alpha,\delta}$; this response encodes the relative alignments of coded mask and sky direction for each of the pointings during an observation, and the energy response. Intensities of celestial source and background models are scaled by coefficients $\Theta_i$ for the source and $\Phi_{i,t}$ for the (time-dependent; identifier $t$) background, which can be determined, e.g., in a fitting procedure minimising the difference between expected (model $m_{(i,j,k)}$) and measured ($d_{(i,j,k)}$) data, expressed in a merit function. As the photons obey the Poissonian counting statistics, our analysis tools (\verb|spimodfit 2.9|, Halloin, 2009) perform maximum likelihood estimations based on the Cash-statistics (negative log-likehihood of Poisson-distributed data; see Cash, 1979) for each energy separately,
\begin{equation}
C(D|\Theta,\Phi) = 2 \sum_k \left[ m_k - d_k \ln m_k \right] \text{, where } \Delta C \propto \chi^2_q,
\label{eq:cash}
\end{equation}
so that the log-likelihood ratio, $\Delta C$, between different models, is proportional to a $\chi^2$-distribution with $q$ degrees of freedom.

Our prime tool of diagnostics for potential data analysis issues or background artefacts is to check and verify the validity of above three assumptions in detail for each particular analysis/study.

\section{Verifications of the Analysis Approach}
\label{sec:verification}

The counts from instrumental background generally dominate over celestial source counts. Therefore it is sufficient here to investigate impacts of the instrumental background, because only this could result in a significant deviation between the model and the measured data. The modelled background is, in general, derived from independent data, i.e from a reference data set $\hat{D}=\hat{d_{(i,j,k)}}$ which differs from the data set $D$ where the source search is made. The intensity and spectrum of cosmic-ray irradiation may undergo significant changes on time scales of hours, e.g. due to solar activity, or if radiation belts are crossed. Hence the background intensity may be different in data sets ($\hat{D},D$). Adopting the background from reference  $\hat{D}$ for observation $D$ may not be perfect. As already pointed out, statistical variations are accounted for in our analysis procedures and algorithm, expressed through Poissonian statistical uncertainty. We analyse potential contamination of the relevant information, and their impacts.

\subsection{Contamination From Transient Background Processes (Time Domain)}

The processes leading to instrumental background may change temporarily, e.g. due to a solar flare, or when the satellite encounters the van Allen radiation belts. A first and simple measure is to exclude the data periods from analysis in which the background model apparently provides an unsatisfactory representation. This can be decided on the basis of the $\chi^2$ contribution of such a period, which would indicate through an exceptionally large value that our background model is not a representative fit during that particular time interval. This represents an allowed \emph{a priori} selection that is independent of the particular source of interest. But such a partial elimination of data also reduces precious source exposure. A more intelligent approach is to investigate if those differences have an impact on analysis results, and how these can be minimised or avoided.

\begin{figure}[!ht]
  \centering
  \includegraphics[width=0.5\linewidth]{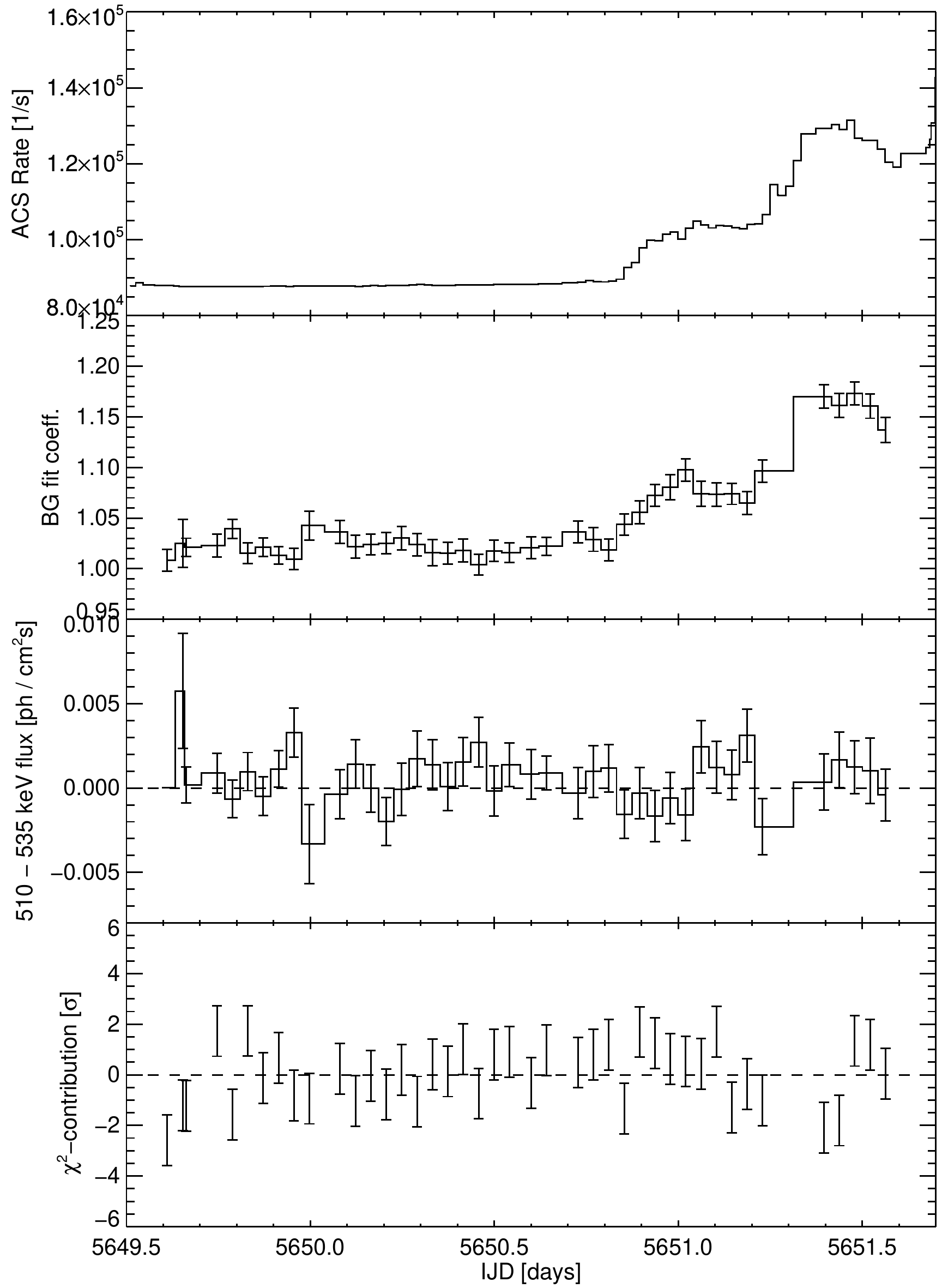}\\
   \includegraphics[width=0.5\linewidth]{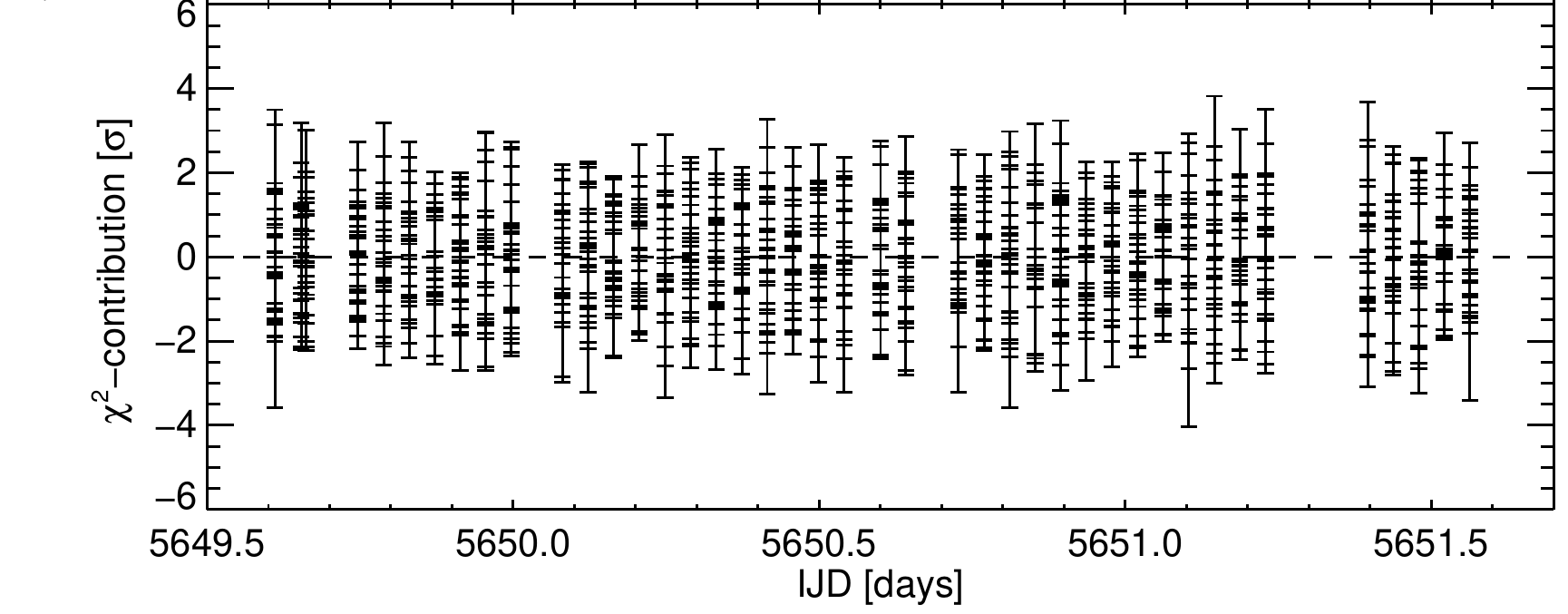}
   \caption{\emph{Upper panels:} INTEGRAL orbit 1555, where background varies. The rate of the BGO anticoincidence system detector counts \emph{(top)} show that increased charge particle irradiation occurs after day 5650.8. The fitted intensities per pointing of the background \emph{(second graph from top)} reflects this background increase, while the celestial signal is found to remain above zero during this entire orbit \emph{(third graph from top)}. The \emph{bottom} graph shows that the quality of the model fitted to measured data remains consistently acceptable during the entire orbit. \emph{Lower panel:} Background scaling and fit quality for all active detectors (only detector 0 was shown in the \emph{(fourth graph from top)}). For all detectors the model fit is adequate, within statistical uncertainties.}
  \label{fig:rev1555_flarecheck}
\end{figure}

The SPI anticoincidence count rate is used to have a measure of the overall intensity of the background, and shows when it undergoes a significant change. The occurrence of a solar flare is represented by count rate increases (Fig.~\ref{fig:rev1555_flarecheck}, \emph{top graph}). Our procedure to re-scale background on short terms appropriately reflects this background intensity change (Fig.~\ref{fig:rev1555_flarecheck}, \emph{second graph from the top}), and the model validity is retained over the entire orbit, and in particular also during the period of increased background intensity (Fig.~\ref{fig:rev1555_flarecheck}, \emph{bottom graph of upper panel}). The intensity of celestial emission is found to remain $\sim$constant (Fig.~\ref{fig:rev1555_flarecheck}, \emph{third graph from the top}), in spite of this variable background. In this case, the background was modeled from data $\hat{D}$ of a set of preceding orbits, fixing the relative detector ratios in the background model to the values derived in $\hat{D}$ data. No exclusion of data is necessary, as our model is not invalidated by the rate increase, and the background variation is not reflected in the celestial-source result. 

\begin{figure}[!ht]
  \centering
  \includegraphics[width=0.5\linewidth]{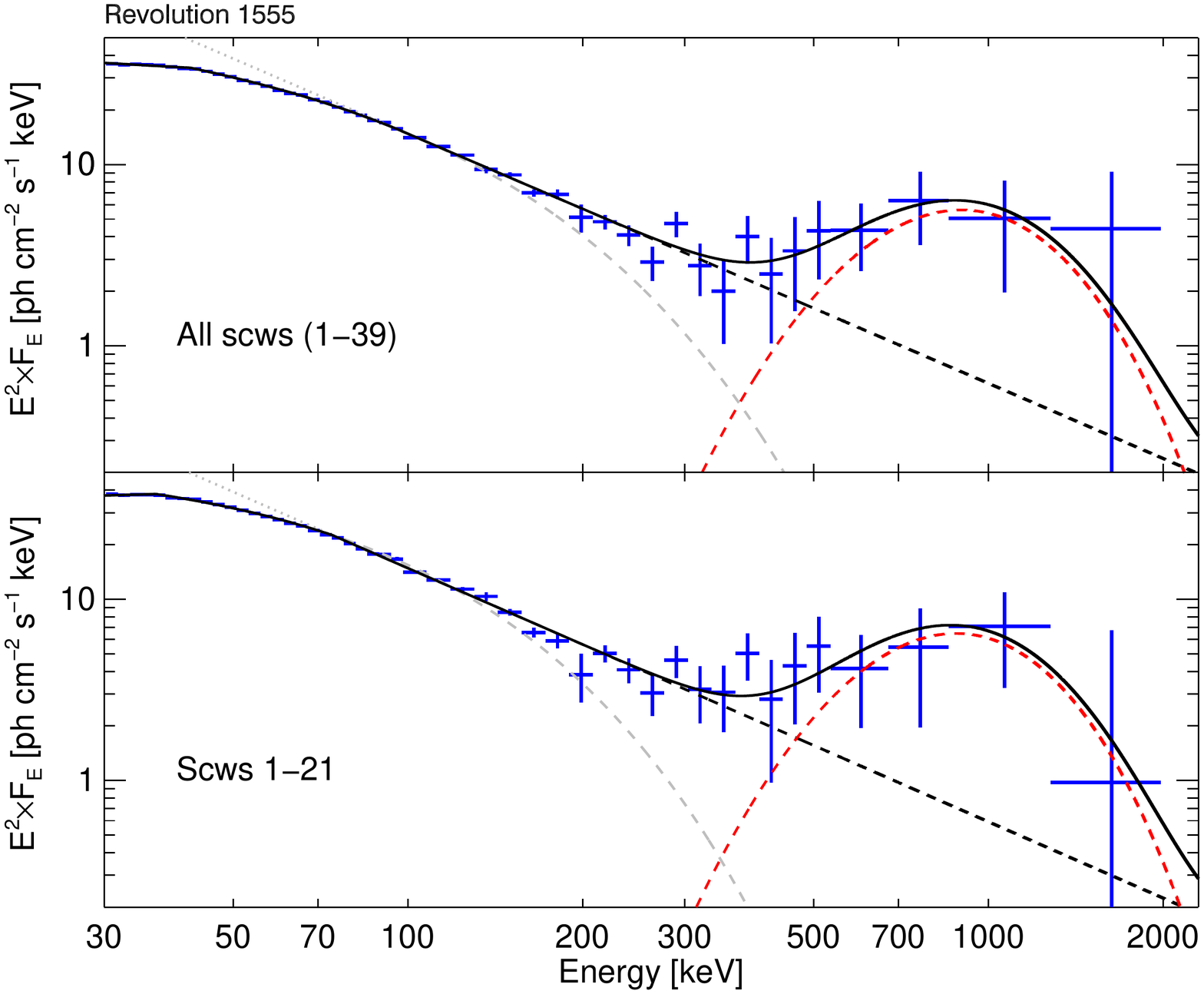}
   \caption{Spectrum of microquasar V404 Cygni during orbit 1555 (``Epoch 1'' in Siegert et al., 2016), for the entire orbit (above), versus the data excluding increased background times (below).}
  \label{fig:v404spec_400}
\end{figure}

This conclusion is further substantiated by comparison (Fig.~\ref{fig:v404spec_400}) of the derived source spectra from data where the increased-background part was excluded with the result from the full data set including the increased background level: The spectral shape and intensity are identical, within the variability expected from the different statistics (only 50\% of the data), leading to an increase of uncertainties.

\subsection{Contamination From Spectral Background Features (Energy Domain)}

\begin{figure}[!ht]
  \centering
  \includegraphics[width=0.4\linewidth]{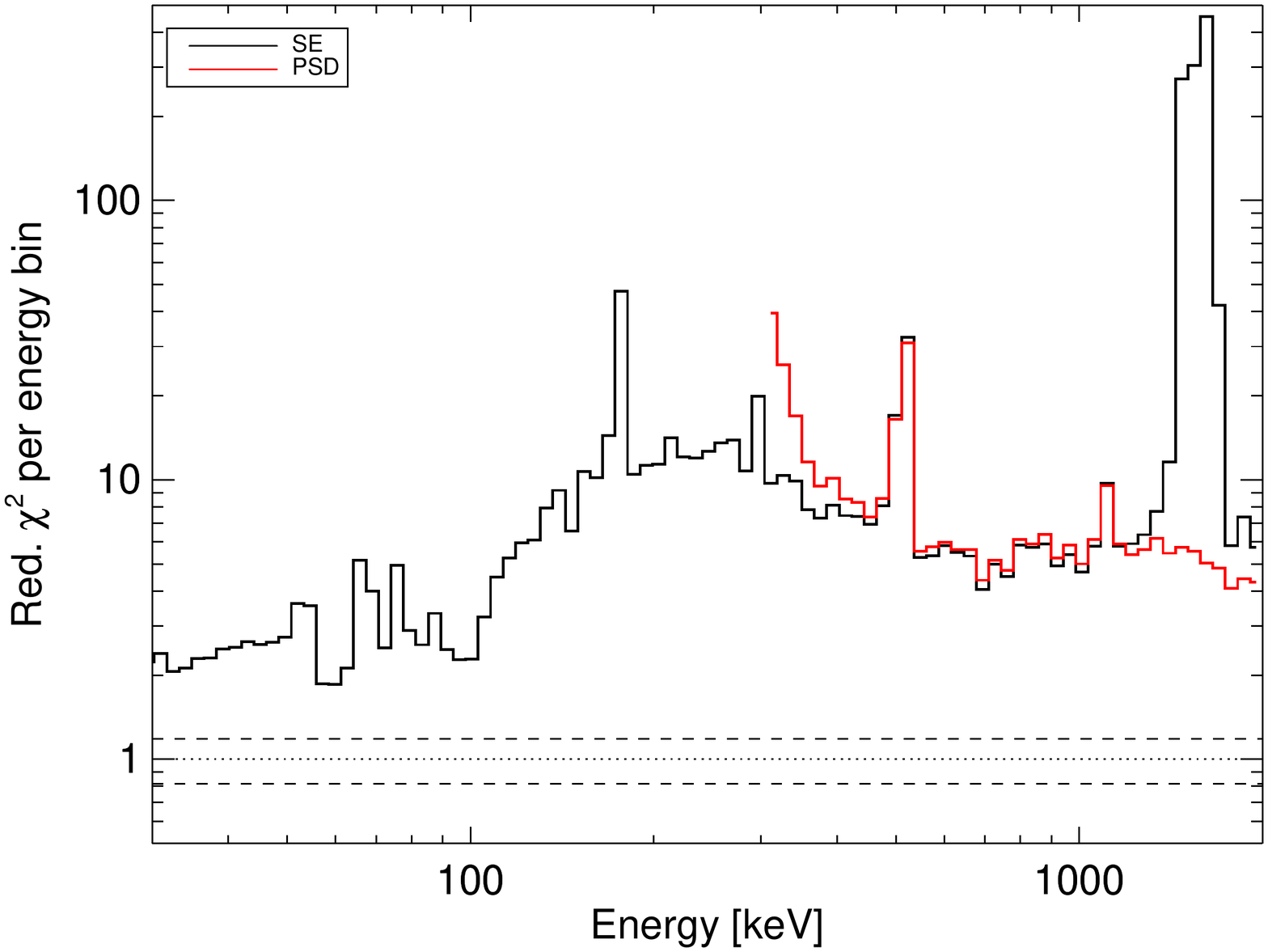}
    \includegraphics[width=0.4\linewidth]{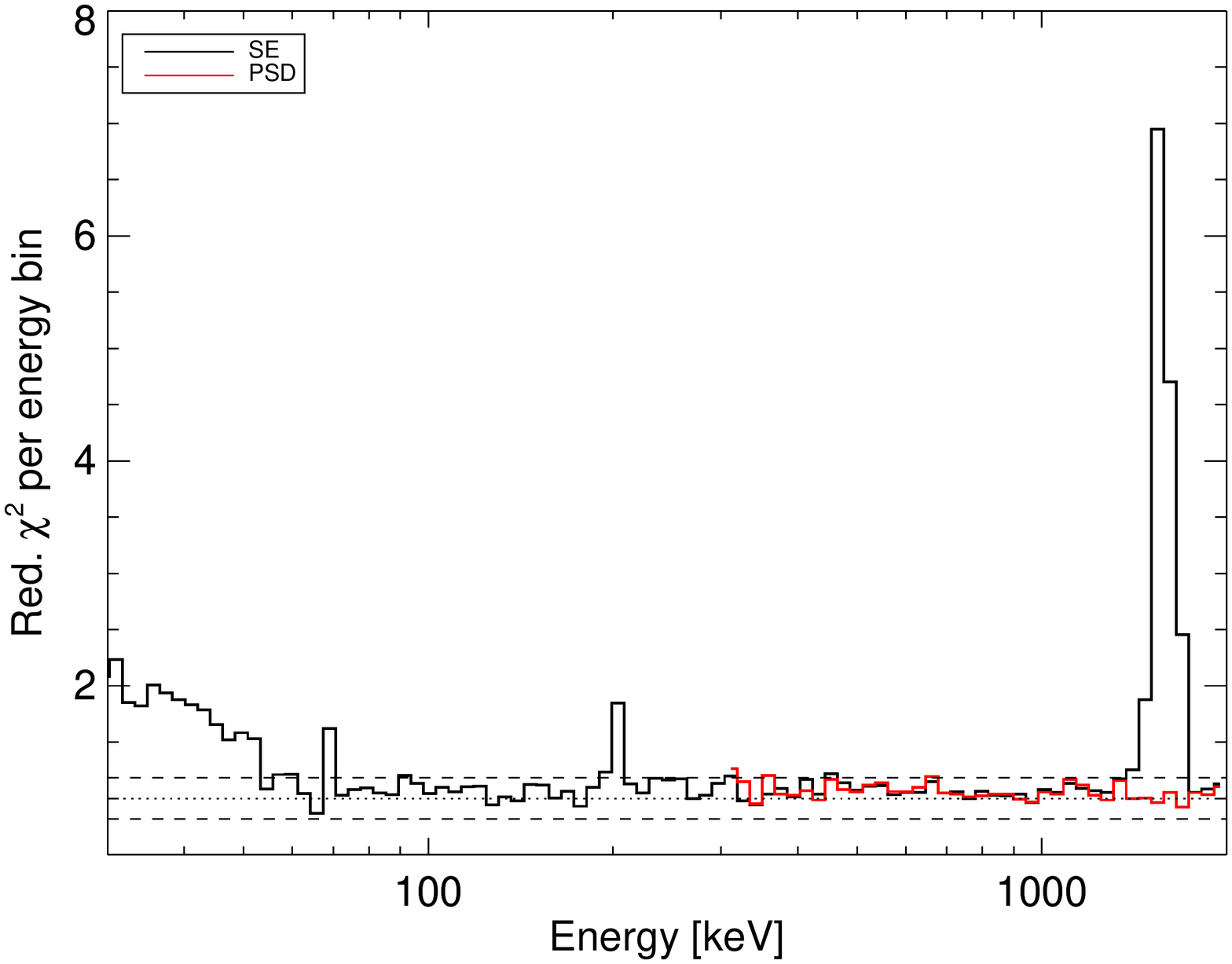}
   \caption{\emph{Left:} Model fit-quality contributions of each energy bin, for a model which has been artificially distorted in its relative detector ratios (see text). The black histogram shows all normal detector triggers, the 'single events' (SE), while the red histogram shows pulse shape selected 'PSD events' (PE) discussed below. \emph{Right:} Fit-quality contributions per energy bin (as above), but now for the real background model as used in our analysis. It is seen that pulse shape selections can recover an adequate fit for the energy range of electronics malfunctions near 1400 keV, while towards lower energies, in particular below 55~keV, fit-quality degrades (see text).}
  \label{fig:chisq_badBgd}
\end{figure}

For high-resolution spectroscopy of celestial sources, instrumental background lines must be properly modeled to exclude their cross talk to the source spectrum. We first illustrate how an inadequate background model with a distorted detector ratio behaves. We take data from orbit 1555 and for two different event selections, `SE' and `PSD' (discussed in detail below). Now we intentionally distort our background model such that relative detector ratios are set to 1.0, i.e. assuming equal background intensities in each detector. This is the assumption adopted in several SPI analysis tools and standard  procedures (SPI User Manual, 2016). With such a background model, the fit quality over energy is seen to be inadequate throughout the entire energy range (Fig.~\ref{fig:chisq_badBgd}, \emph{lefthand panel}). Additionally, spectral variation resembles instrumental lines. For comparison, the fit quality of our \emph{source + background} model as applied in our analysis (Siegert et al., 2016) is shown in the \emph{righthand panel} of Fig.~\ref{fig:chisq_badBgd}, for the same data. The fit is now acceptable (dashed lines) across much of the energy range. But apparently, there still are inadequacies in the fit, most strikingly near 1.5 MeV for the `SE' selection. These fit imperfections are plausibly attributed to spurious behaviour of electronics  that has been found early in the INTEGRAL mission in the 1350-1700 keV energy band, and to imperfections of the SPI low-energy spectral response. The latter is particularly apparent e.g. in the 198 keV line from a cascade event (Naya et al., 1996), and can probably only be addressed by more accurate background modeling in finer energy bins. 

\begin{figure}[!ht]
  \centering
  \includegraphics[width=0.4\linewidth]{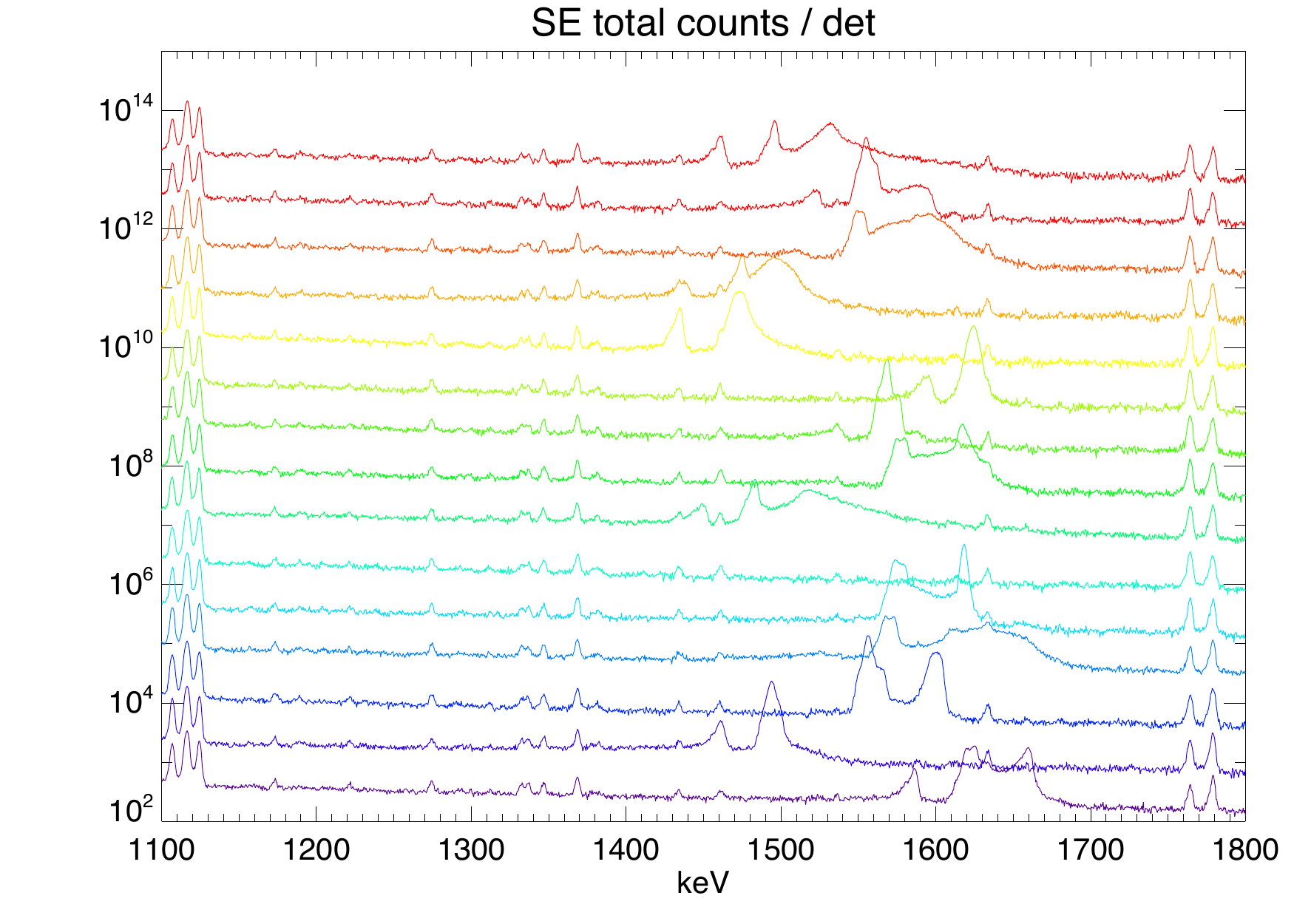}\\
    \includegraphics[width=0.4\linewidth]{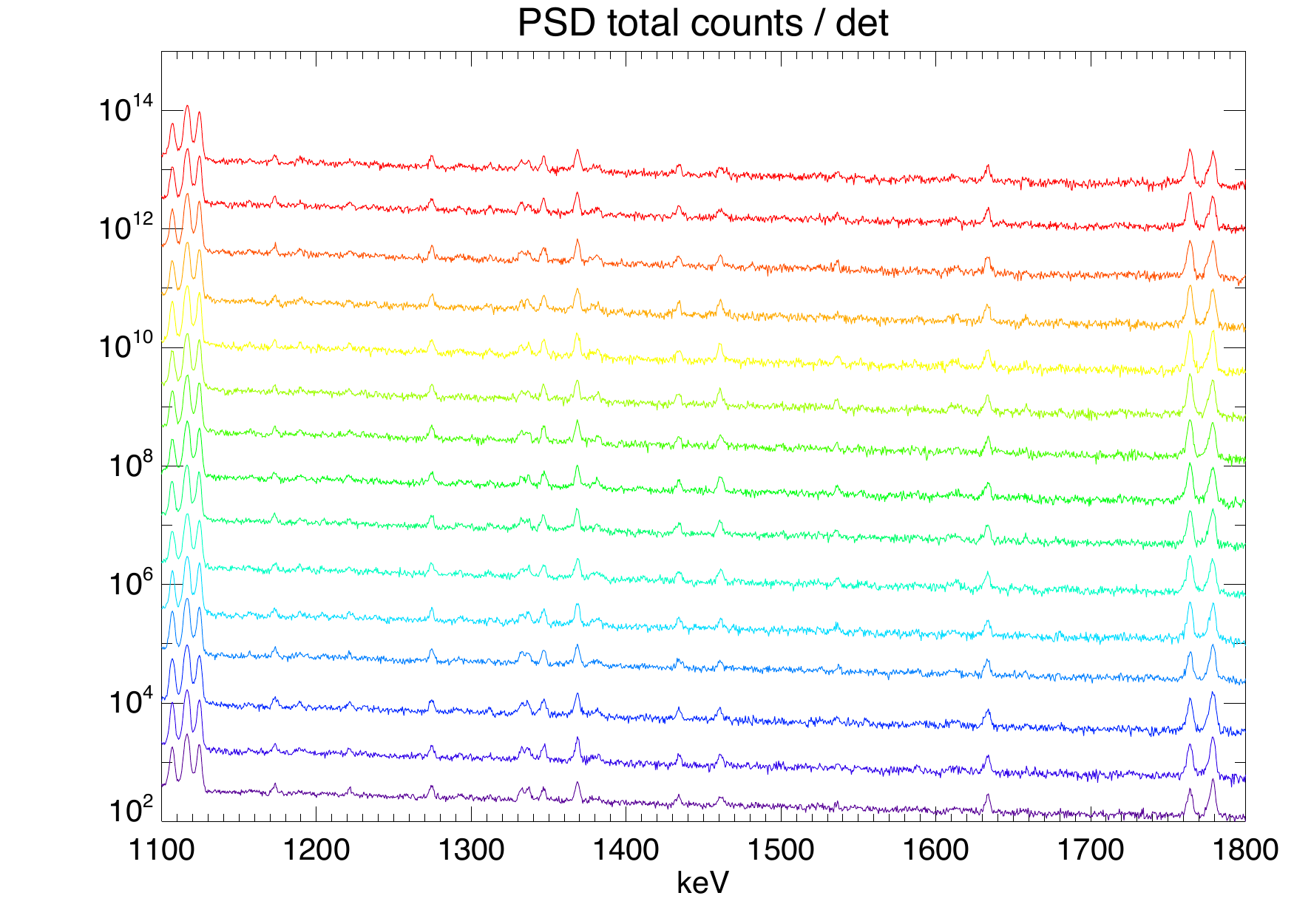}
   \caption{Event rates for all events ('SE'; above) and pulse shape selected events 'PE' (below), for orbit 1557. The different lines are spectra for different SPI detectors, offset with respect to each other; the abscissa units are valid for detector 0, the lowest line shown.}
  \label{fig:DetSpec_SE_PE}
\end{figure}

As for the electronics artefacts: The SPI electronics features an additional analogue chain that evaluates pulse shape information. Depending on settings, this electronics handles events with energies above 400 keV, and provides an additional event flag, PSD, identifying consistency of the signal pulse shape with expectations for an event originating from a gamma-ray photon interaction. It also has been found that the electronics features in the 1350-1700 keV energy band can be eliminated, while otherwise the spectral response is identical to the PSD-unselected events at all other energies (see Fig.\ref{fig:DetSpec_SE_PE}). `SE' therefore is a data set ignoring pulse shape information, while `PSD' (`PE') data make use of the rejection of events with pulse shapes that are not recognised as due to gamma-ray triggers. 

\begin{figure}[!ht]
  \centering
  \includegraphics[width=0.4\linewidth]{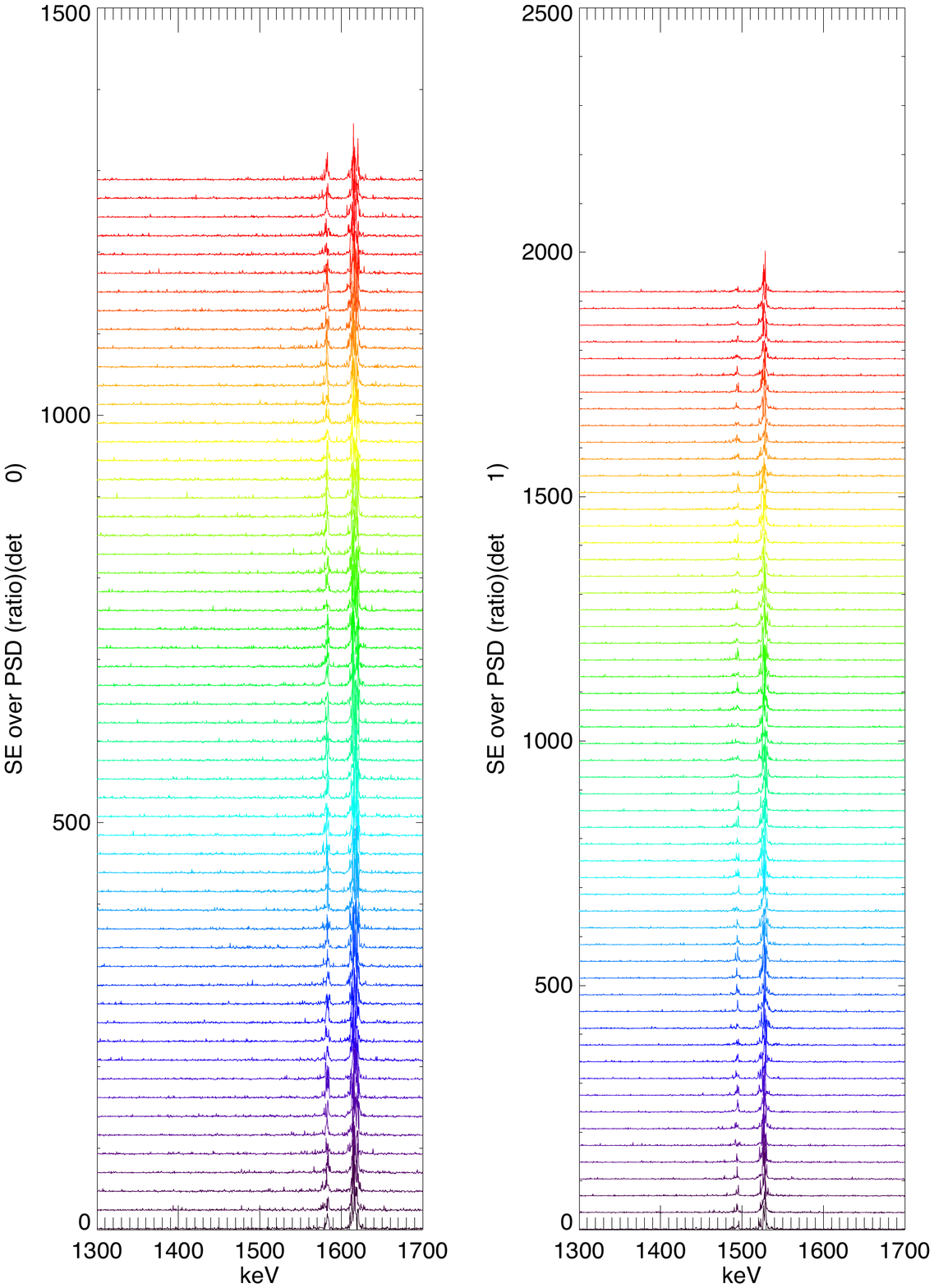}
   \caption{Spectral features from electronics malfunctions, shown as a function of time during one orbit (no. 200), and for two different detectors: The spectral behaviour per detector appears rather constant during such time interval.}
  \label{fig:DetSpec_rev200}
\end{figure}

The reasons for this drastic degradation of the model quality in the 1350-1700 keV energy band are also clear from Fig.~\ref{fig:DetSpec_SE_PE}: The electronics artefacts occur at different energies for each of the individual camera detectors. Therefore the detector-to-detector ratio is drastically different for those events. Any variation of this contamination now leads to different detector-to-detector ratios between model and observations data sets for the `SE' data. Again, a careful improvement of background modelling may be employed: Fig.\ref{fig:DetSpec_rev200} shows that the electronics artefact signature per each detector appears to remain rather constant during at least one orbit. Therefore, in a detector-per-detector spectral background model, the electronics artefacts are identical in appearance to instrumental lines, and thus can be handled identically. So, in a high-spectrally resolved background model, it should be possible to recover the good background fit also for `SE' data in Fig.~\ref{fig:chisq_badBgd}. This remains to be shown for 1350-1700 keV data, however. In V404 Cygni data analysis, broader energy bins have been used for background modelling, and therefore the selected subset of `PE' data have been used across this energy range of electronics background. 

\begin{figure}[!ht]
  \centering
  \includegraphics[width=0.33\linewidth]{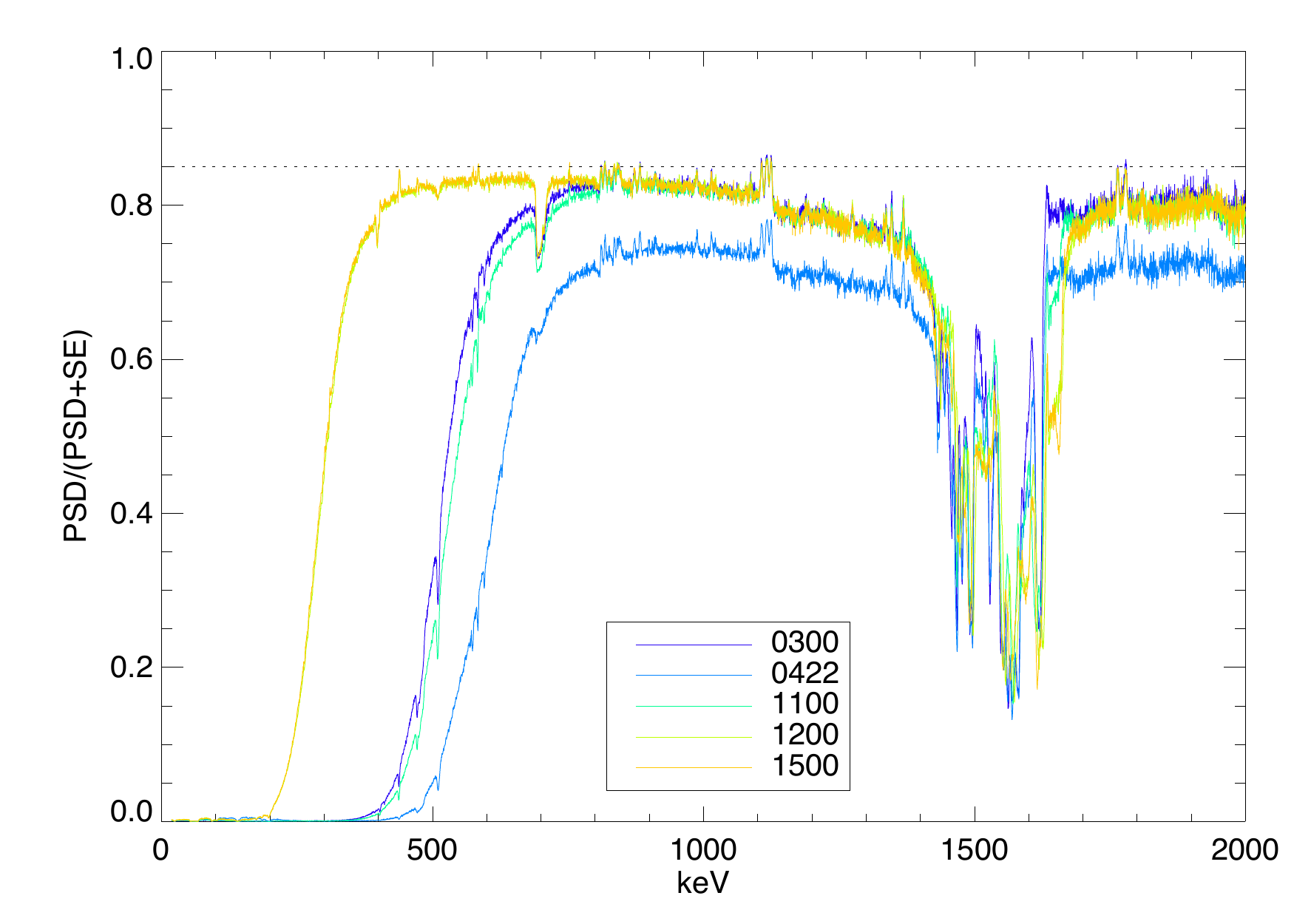}
    \includegraphics[width=0.3\linewidth]{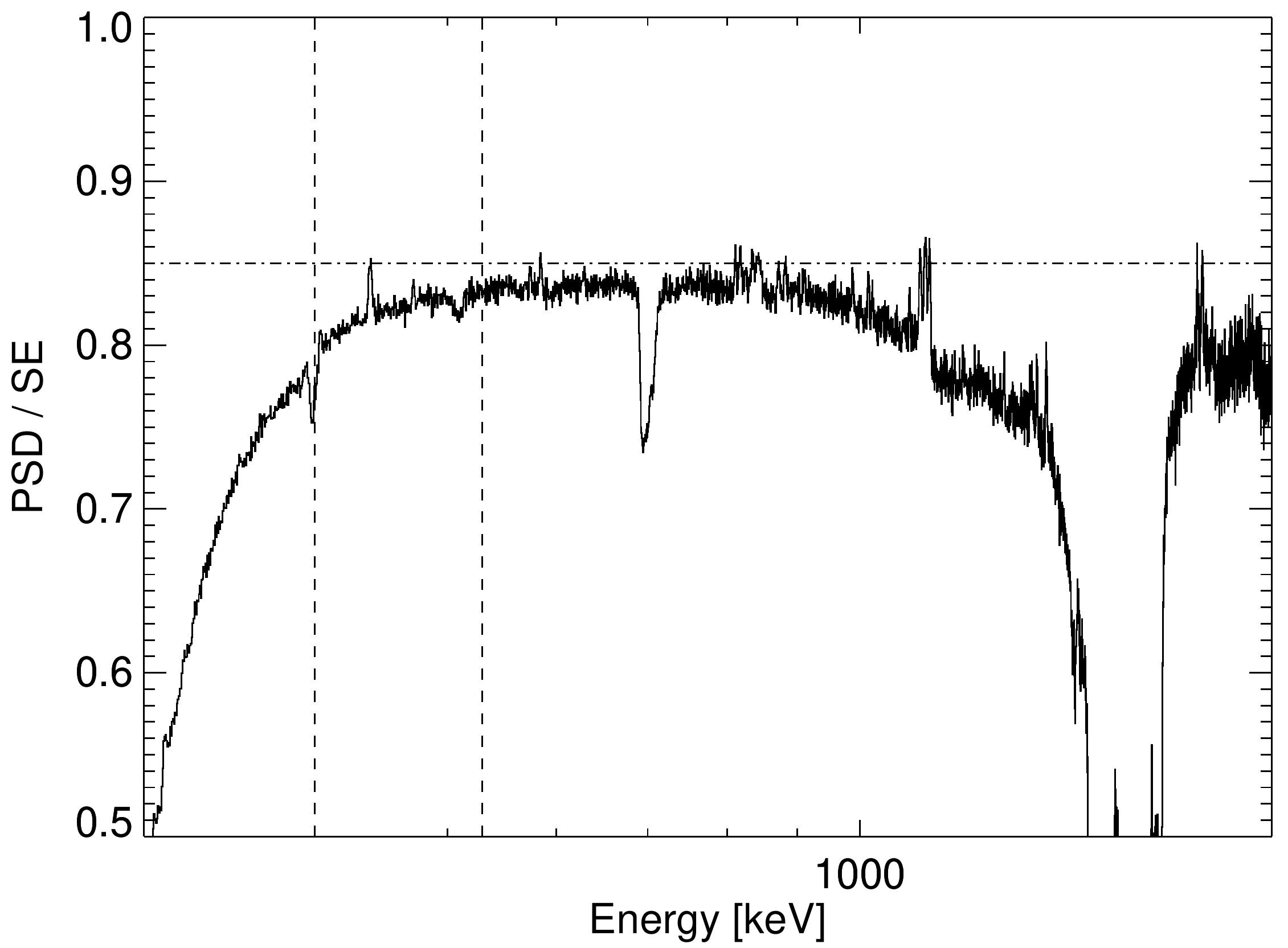}
      \includegraphics[width=0.3\linewidth]{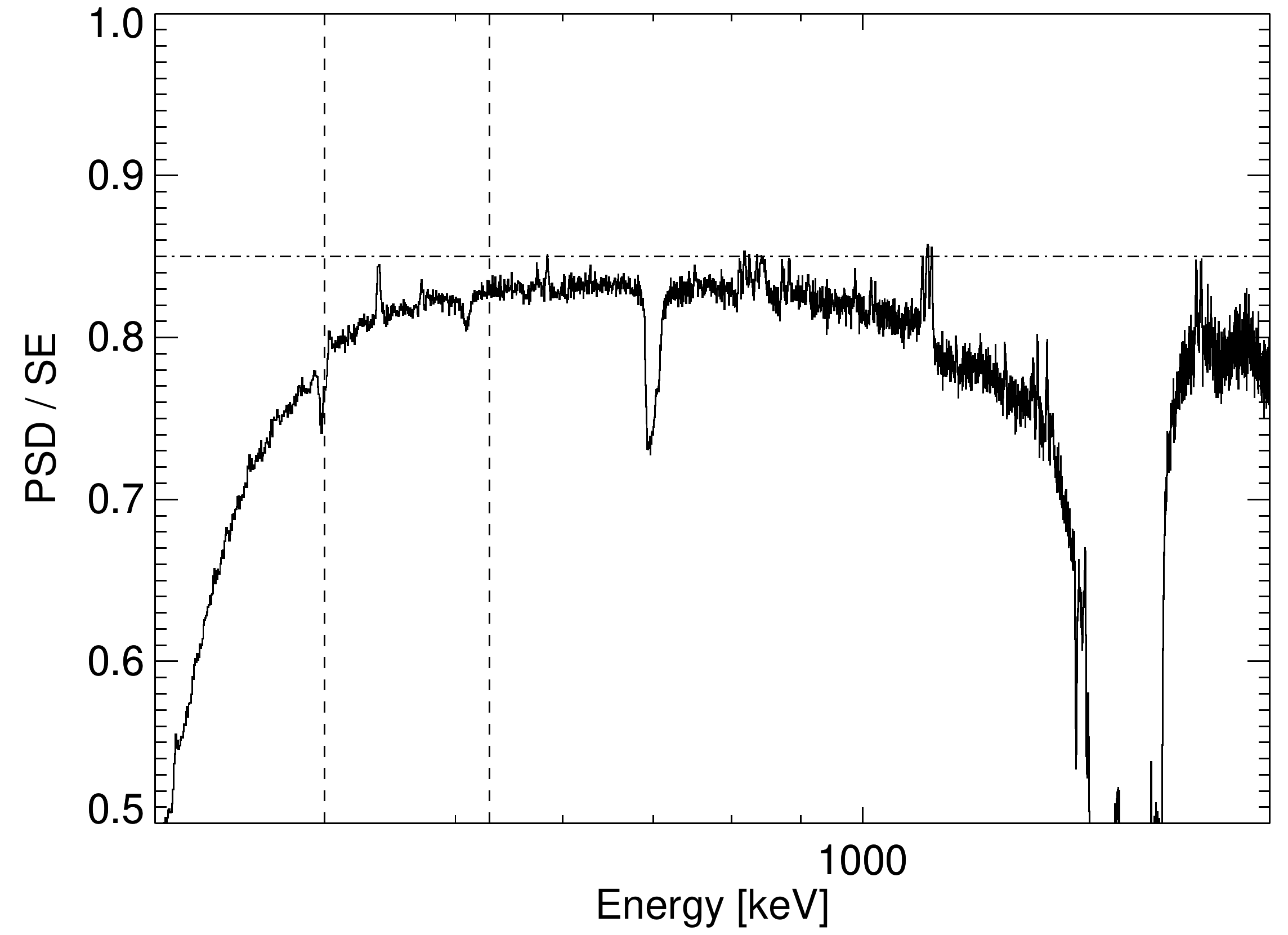}
   \caption{Spectral behaviour of the pule shape selection, shown for five different orbits during the INTEGRAL mission (left), and for the V404 Cygni flare (orbit 1557 (center) and for its background reference (orbit 1552, right).}
  \label{fig:PE-SE-ratio}
\end{figure}

Since pulse shape information is only available in a limited energy range as given by the properties and settings of the PSD electronics, the combination of unselected SE data with selected PE data must apply an appropriate selection efficiency factor for PE. Fig.~\ref{fig:PE-SE-ratio} shows how this correction factor might be extracted from the data, determining the PE/SE event ratio of events that satisfy the gamma-ray pulse shape criteria (PE) versus the unselected events. Evidently, operational changes affect the PSD electronics response at low energies up to $\sim$800 keV (top Fig. 9), but the spectral signatures remain at same energies throughout, in particular in the 1350-1700 keV band of the electronics artefacts. Between data used for background modelling and for V404 Cygni analysis, the event ratio remained unchanged, and no systematics from changes of PSD electronics are expected (Fig.~\ref{fig:PE-SE-ratio}, \emph{rightmost two plots}). 
The ratio of events always remains significantly below 1.0, and thus on average a fraction $\sim$20\% of gamma-ray events is suppressed by this selection process, also reducing data from the celestial sources. Assuming that the 1350-1700 keV energy range shows us events that only are due to background, one may estimate a selection correction factor from the highest values obtained at other energies, interpreting all line-like features as background-only events. This optimistic estimate reaches up to the 85\% efficiency that had been determined from electronics measurement in the laboratory prior to the INTEGRAL launch. More realistically, there is a smooth trend in selection efficiency, with values of $\sim$0.75 at 400 and 1200 keV and $\sim$0.82 in between.
   
Note that, when aiming for higher statistical precision, often broader energy bins are chosen for background modeling, which then include multiple and different background origins. In this case, the background detector pattern as modeled from $\hat{D}$ will, in general, differ from the one in the source data set $D$. Its poorer fit to the current real background detector pattern hence will leave some systematically deviant residuals. These could partially correlate (or anti-correlate) with a detector-variable pattern as it corresponds to the celestial source. The inspection of the celestial source spectrum in detail and its comparison to the instrumental background spectrum will again be a valuable diagnostics for such impacts of inadequate background modelling. We demonstrated this for $^{56}$Ni line spectroscopy in the case of SN2014J (see Appendix of Diehl et al., 2014).

\begin{figure}[!ht]
  \centering
  \includegraphics[width=0.4\linewidth]{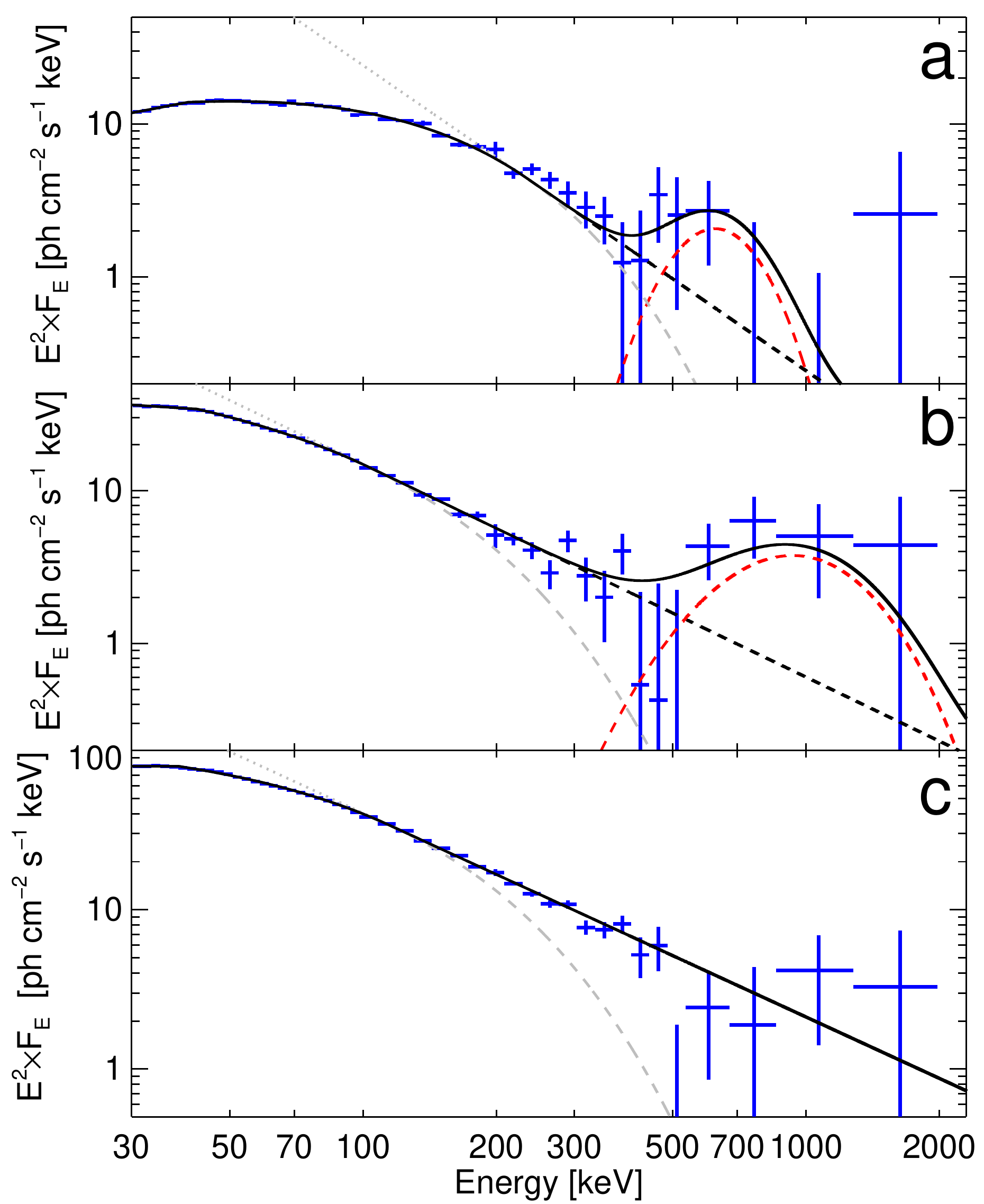}
   \caption{Spectra of V404 Cygni in three epochs, as shown in Fig.~1 of Siegert et al. (2016), but here with pulse shape selections applied already at lower energies $E>400$~keV, as suggested by Roques et al. (2016).}
  \label{fig:Spec_v404_3_400}
\end{figure}

Towards a similar consistency check of energy spectra, we also compare the spectra derived for V404 Cygni in the three different epochs, as derived for different selection windows for using pulse shape information: These spectra (Fig.~\ref{fig:Spec_v404_3_400}) have been derived using pulse shape selections for all event energies above 400 keV, and compare to the spectra of our paper, Fig. 1, which were derived using pulse shape selection above 530 keV. No significant differences arise for the first two epochs. Spectral fit parameters are consistent. The third epoch still shows a highly significant high-energy excess, however the positronium edge seems to be smoothed out.

We point out systematic uncertainty in the selection bias factor (efficiency), and our general uncertainty on the need, justification, and usefulness of pulse shape selections beyond the 1350-1700 keV interval.

\begin{figure}[!ht]
  \centering
  \includegraphics[width=0.8\linewidth]{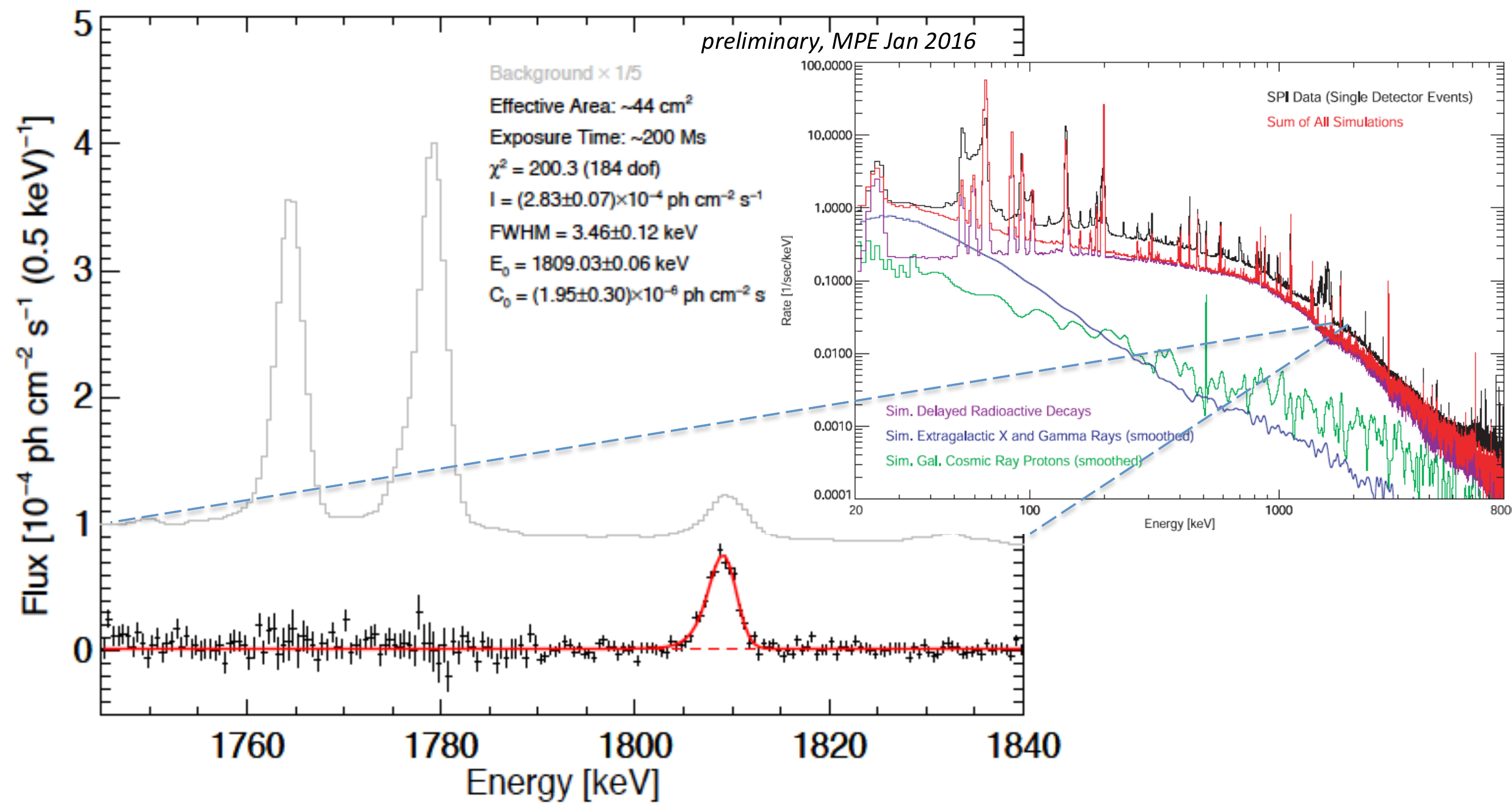}
   \caption{Spectrum of Galactic emission as recognised in model fitting, showing the $^{26}$Al line at 1809 keV. The down-scaled raw background spectrum is shown as the grey histogram. The graph on the righthand side shows the entire SPI background spectrum, indicating the region detailed on the left graph. It is evident that instrumental background lines are not "shining through" the spectrum derived for the sky.}
  \label{fig:26Al-sim}
\end{figure}

We also analyse another data set collected from all events during 13 years of INTEGRAL observations for the range above the 1350-1700 keV range of electronics malfunction contamination, and investigated if any background model issues may appear. Strong background lines at 1764, 1779, and 1810 keV (Fig.~\ref{fig:26Al-sim}, \emph{top histogram}) can be seen just above the energies with the suspicious electronics artefacts (\emph{inset} in Fig.~\ref{fig:26Al-sim}, showing simulated versus measured data, and the electronics artefacts not appearing in simulations based on instrument and detector physics). 

In the spectrum extracted for the Galaxy, only the $^{26}$Al line shows up, and all instrumental background lines are properly identified and thus absent from the celestial source spectrum. We take this as additional support that electronics artefact events do not significantly contaminate the measurements at energies beyond the 1350-1700 keV range, and unselected SE events with their $\sim$15-30\% higher exposure may be used.

\begin{figure}[!ht]
  \centering
  \includegraphics[width=0.7\linewidth]{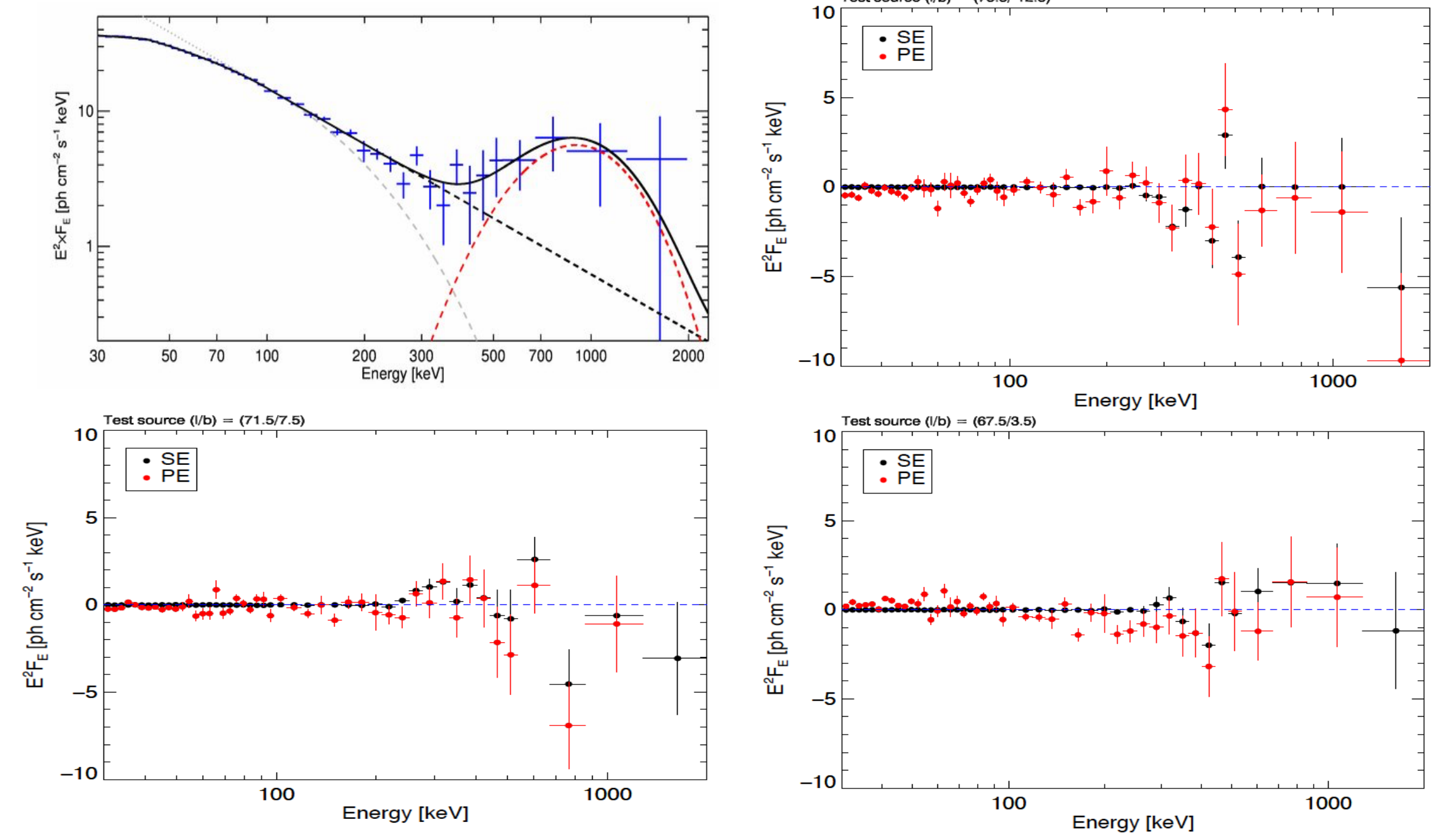}
   \caption{Spectra for four different positions in the observing field around the microquasar V404 Cygni. The spectrum of V404 Cygni (top left) shows clear high-energy emission, while other locations in the region near V404 Cygni do not show any systematic or significant emission in SPI data.}
  \label{fig:specs_4Src}
\end{figure}

\subsection{Contamination From Unrecognised Background Features (Sky Domain)}

Another powerful tool of diagnostics is to check how sharp the correlation of the source signal with the expected complex coding pattern depends on where the candidate source is located: If statistical fluctuations of the data would lead to a false identification of measured counts with a celestial source, those false identifications should occur not just for one particular source on the sky, but also for sources in their vicinity within the field of view of the instrument. Also for systematic inadequacies of the background model, appearance of artefacts for different source directions within the field of view would be expected. By definition, background events are ignorant about the coding pattern imposed for a celestial source by the coded mask and the dithering observations. Therefore, if a systematic \emph{source signal} is found for more than one position on the sky, then the source could either be extended/diffuse, or the signal could be the result of background inadequacies. Also, if positive and negative sources (i.e. sources which appear significant but have negative flux values) appear within the observed field, the cause of this often is an inadequate background model: Residual counts not well represented by the background model may be captured by the only other degree of freedom of the fit, i.e. an \emph{(artificial) source} in the sky. But its statistical origins translate into apparently negative flux distributions. 

For such diagnostics, we perform analyses where the adopted position of the candidate source is varied systematically across the inner instrument field of view. If a signal is present in larger regions of the sky, it may be suspected to be spurious and induced by background model imperfections. Fig.~\ref{fig:specs_4Src} shows a set of spectra for different positions of candidate sources, from the data for the flaring period of V404 Cygni. Significant emission shows up only for V404 Cygni, and not for other positions. Spectra for other regions are consistent with zero fluxes, within statistical uncertainty. Even though at energies towards 1 MeV the uncertainties increase, and lead to outliers within $\pm$(1-2)$\sigma$ of statistical uncertainty. Only V404 Cygni shows a consistent bump around 1 MeV, which we associate with pair plasma and positron annihilation. There are no indications of background inadequacies or contamination, from other pivot points on the sky showing no similar excess signals in the few hundred to 1 MeV region. 

\section{Additional References}
Cash, W., ApJ 228, 939. (1979) \hfil\break
Diehl R., et al., Science, 345, 6201, 1162 (2014) \hfil\break
Diehl R., Rep. Prog. Phys., 76, 026301 (2013) \hfil\break
Haloin, H., spimodfit Explanatory Guide and Users Manual, (2009) \hfil\break
Kretschmer K. A., Dissertation, TU M\"unchen (2012) \hfil\break
Naya J. E., et al., Nucl. Instr. Meth., A368, 832 (1996) \hfil\break
Siegert T. et al., Nature, 531, 341 (2016); this work \hfil\break
Siegert T., Master Thesis, TU M\"unchen (2013) \hfil\break
SPI User Manual, ESA AO-14 Documentation (2016)

\end{document}